\documentclass[a4paper]{article}
\usepackage{graphics}
\newenvironment{myfig2}[1]{\begin{figure}[#1] 
  \begin{center}
    \begin{tabular}{p{.465\textwidth}p{.01\textwidth}p{.465\textwidth}} }
    {\end{tabular}
  \end{center} 
 \end{figure}
}

\begin{document}


\begin{titlepage}
\hspace*{-5mm}Available on CMS information server
\hfill {\Large\bf CMS NOTE 1998/073}

\begin{center}
\includegraphics*{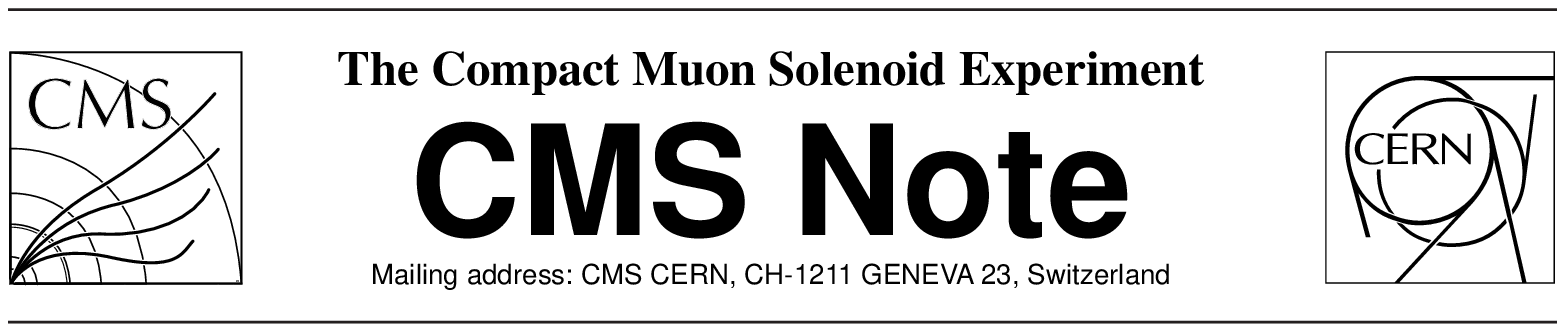} 
\end{center}
\begin{flushright}{\bf November 10, 1998}\end{flushright}

\vspace*{15mm}

\begin{center}
{\Large\bf Search for SUSY in (leptons +) Jets + E$_T^{miss}$ Final States}
\end{center}

\vspace{15mm}

  \begin{center}
    S.~Abdullin~$^{a)}$, F.~Charles~$^{b)}$,\\
     Groupe de Recherche en Physique des Hautes Energies\\
      Universit\'{e} de Haute Alsace, 61 rue A.Camus 68093 Mulhouse, France
  \end{center}

\vspace{20mm}

  \begin{abstract}
We study the observability of the strongly interacting squarks and gluinos 
in CMS. Classical E$_{T}^{miss}$ + jets final state as well as
a number of additional multilepton signatures  
(0 leptons, 1 lepton, 2 leptons of the same sign, 2 leptons of the opposite
 sign and 3 leptons) are investigated .
 The detection of these sparticles relies on the observation of an
excess of events over Standard Model background expectations. 
The study is made in the framework of a minimal
SU(5) mSUGRA model as a function of m$_{0}$, m$_{1/2}$ for 4 sets of
model parameters $:$ tan$\beta$ = 2 or 35 and sign($\mu$) = $\pm$1 
and for fixed value of A$_{0}$ = 0. 
The CMS detector response 
is modelled using CMSJET 4.51 fast MC code (non-GEANT). The results obtained
are presented as 5 $\sigma$ detection contours  in the
 m$_{0}$, m$_{1/2}$ planes 
 and with optimized selection cuts in  various regions of the parameter space.
The result of these investigations is that
with integrated luminosity L=10$^{5}$  pb$^{-1}$ the squark and gluino
 mass reach is about 2.5 TeV
and covers most of the interesting parts of parameter space according to
neutralino relic density expectations. The influence of signal and
 background cross-section uncertainties on the reach contours is estimated. 
 The effect of pile-up on signal and
background is also discussed. This effect 
is found to be insignificant for E$_{T}^{miss}$ and single lepton signatures,
whilst only a minor deterioration is seen for multilepton final states.
  \end{abstract}

\vspace{20mm}
\hspace*{-8mm} \hrulefill \hspace{80mm} \hfill \\
$^{a)}$ On leave from ITEP, Moscow, Russia.
  Email: adullin@mail.cern.ch\\
$^{b)}$ Email: charles@in2p3.fr  

\end{titlepage}
\pagenumbering{arabic}
\setcounter{page}{1}
\section{Introduction}

One of the main purposes of the LHC collider is to search for the physics 
beyond the Standard Model (SM). One of the direction of this search is
a possible discovery of  superpartners of ordinary particles as expected in 
Supersymmetric extensions of SM (SUSY). 
SUSY, if it exists, is expected to reveal itself at LHC via excess of
 (multilepton +) multijet + E$_{T}^{miss}$ final states
 compared to Standard Model (SM) expectations \cite{reach_papers}. 

The main goal of this study is to evaluate the potential of the CMS detector
\cite{CMS}
to find evidence for SUSY. It deals, first with the squarks and 
gluino mass reach, as the production cross-section of these strongly 
interacting sparticles (pair production or in association with charginos and
neutralinos) dominates the total 
SUSY cross-section over a wide region of the parameter space.
In our previous study concerning maximal reach in mSUGRA 
for low tan$\beta$ \cite{mynote} the SM background was somewhat  
underestimated; furthermore the signal form squarks and gluinos
 was taken into account, without associated production of squarks
 and gluino with electroweak sparticles.
 Besides the lepton isolation requirements at the tracker
level were somewhat unrealistic (too small cone size R$_{isol}$=0.1)
 allowing electrons to be ``isolated'' in
jets. Here we make a substantial revision of the results previously obtained
and extend our search to the domain of large values of tan$\beta$.
The effect of event pile-up  on the possible mSUGRA reach is also investigated.

The paper is organized as follows.
We discuss the specific SUSY model employed in section 2.
In section 3 the simulation procedure issues 
are presented. Comparison of same signal and background distributions 
is shown in section 4. Section 5 describes the cuts optimization procedure 
allowing to adjust the cuts proposed to the condition in various domains
of the model parameter space.
The main results of our study are presented in section 6 where we also discuss
the stability of the reach contours versus various sources of uncertainty.
The conclusions  are given in section 7.

\section{Model employed}

The large number of SUSY parameters even in the framework of
Minimal extension of the SM (MSSM) makes it difficult to evaluate the
general reach. So, for this study
we restrict ourselves  at present to the mSUGRA-MSSM model. This model
evolves from MSSM, using Grand Unification Theory (GUT) assumptions
(see more details in e.g. \cite{msugra}).
In fact, it is a representative model, especially in case of inclusive studies
and reach limits expressed in terms of squark and gluino mass which 
do not depend critically on the specific choice of branchings and mass 
values as will be indirectly shown by the results of this work. 
 
The mSUGRA model contains only five free parameters $:$
\begin{itemize}
 \item a common gaugino mass ($m_{1/2}$) ;
 \item a common scalar mass ($m_{0}$);
 \item a common trilinear interaction amongst the scalars ($A_{0}$);
 \item the ratio of the vacuum expectation values of the Higgs fields that
       couple to $T_{3} = 1/2 $  and $T_{3} = -1/2$ fermions ( tan$\beta$);
 \item a Higgsino mixing parameter $\mu$ which enters only through its
   sign ($sign (\mu)$).
\end{itemize}

 For a given choice of model parameters all
 the masses and couplings, thus production cross sections and branching ratios
are fixed. At a later stage it can be generalized to the MSSM in which no such
constraining  relations exist.

The mass of the lightest SUSY particle (LSP) which is 
$\tilde{\chi}_{1}^{0}$ in the R-parity conserving mSUGRA equals approximately
$\sim$ 0.5 of $\tilde{\chi}_{2}^{0}$. The mass of lightest chargino
 $\tilde{\chi}_{1}^{\pm}$ is almost
the same as that of  $\tilde{\chi}_{2}^{0}$. Isomass contours of 
$\tilde{\chi}_{1,2}^{0}$ and  $\tilde{\chi}_{1}^{\pm}$ and gluino behave
 gaugino-like, i.e. depend mainly on $m_{1/2}$. Masses of sleptons and squarks
depend on both $m_{0}$ and $m_{1/2}$.

Masses of squarks (especially of the first generation), gluino, charginos and
neutralinos depend only weakly on tan$\beta$,
 A$_{0}$ or sign($\mu$) parameters. 
Masses of sleptons, stop and sbottom
have some dependence on these mSUGRA
parameters, whilst masses of Higgs bosons depend significantly on 
tan$\beta$ (see some examples e.g. in \cite{large_tan}), the mass
of lightest scalar Higgs increases with tan$\beta$ 
and depends also on sign($\mu$), whilst 
masses of the heavy Higgses decreases dramatically with  tan$\beta$.

Since masses, branchings, cross-sections vary most rapidly with 
 m$_0$, m$_{1/2}$, it is natural to follow the commonly used way of 
presenting mSUGRA data as a function of these two parameters for 
different fixed values of  tan$\beta$ and sign$(\mu$). The A$_0$ parameter
 is usually set to zero, since its variation has small effect on the results.
Figs.1,2 
show isomass contours of SUSY particles for some particular 
choice of mSUGRA parameters, namely $:$ tan$\beta$ = 2, A$_{0}$ = 0
and $\mu$  $<$ 0 just to have some idea about the characteristic values
and behaviour of the masses versus  m$_0$ and m$_{1/2}$.
 The shaded regions along the axes denotes theoretically
(TH) and up to now experimentally (EX) excluded regions of the model 
parameter space. The data concerning these regions were taken from 
\cite{ex_regions} .

The MSSM establishes the relation between the top mass and tan$\beta$
 (see e.g. \cite{Boer}) $:$
\begin{center} m$^{2}_{t}$ = 4$\pi$Y$_{t}v^{2}$
                  $\frac{tan^{2}\beta}{1 + tan^{2}\beta}$  \hspace{4cm}(1)
\end{center}
Here Y$_{t}$ is top Yukava coupling.
For low tan$\beta$ the top Yukava coupling can be derived from known gauge
couplings alone, which leads to the tan$\beta$ value of 1.6$\pm$0.3 for
m$_{t}$ = 175$\pm$6 GeV. Taking into account behaviour of
bottom and $\tau$ Yukava couplings at large tan$\beta$ (see discussion e.g.
in \cite{large_tan}), it seems to be possible to find second solution  
of eq.(1) with tan$\beta$ $\approx$ 33$\pm$3 (see also Fig.6 in \cite{Boer}).
So, from now on we will consider mainly dependence of various mSUGRA
 observables and will present our results for 4 sets of
tan$\beta$ and $\mu$ parameters, keeping  A$_0$=0, see Table.1.
\vspace*{-5mm}
\begin{myfig2}{hbtp}
\vspace*{0mm}
\hspace*{-3mm}\resizebox{.465\textwidth}{10cm}
                      {\includegraphics{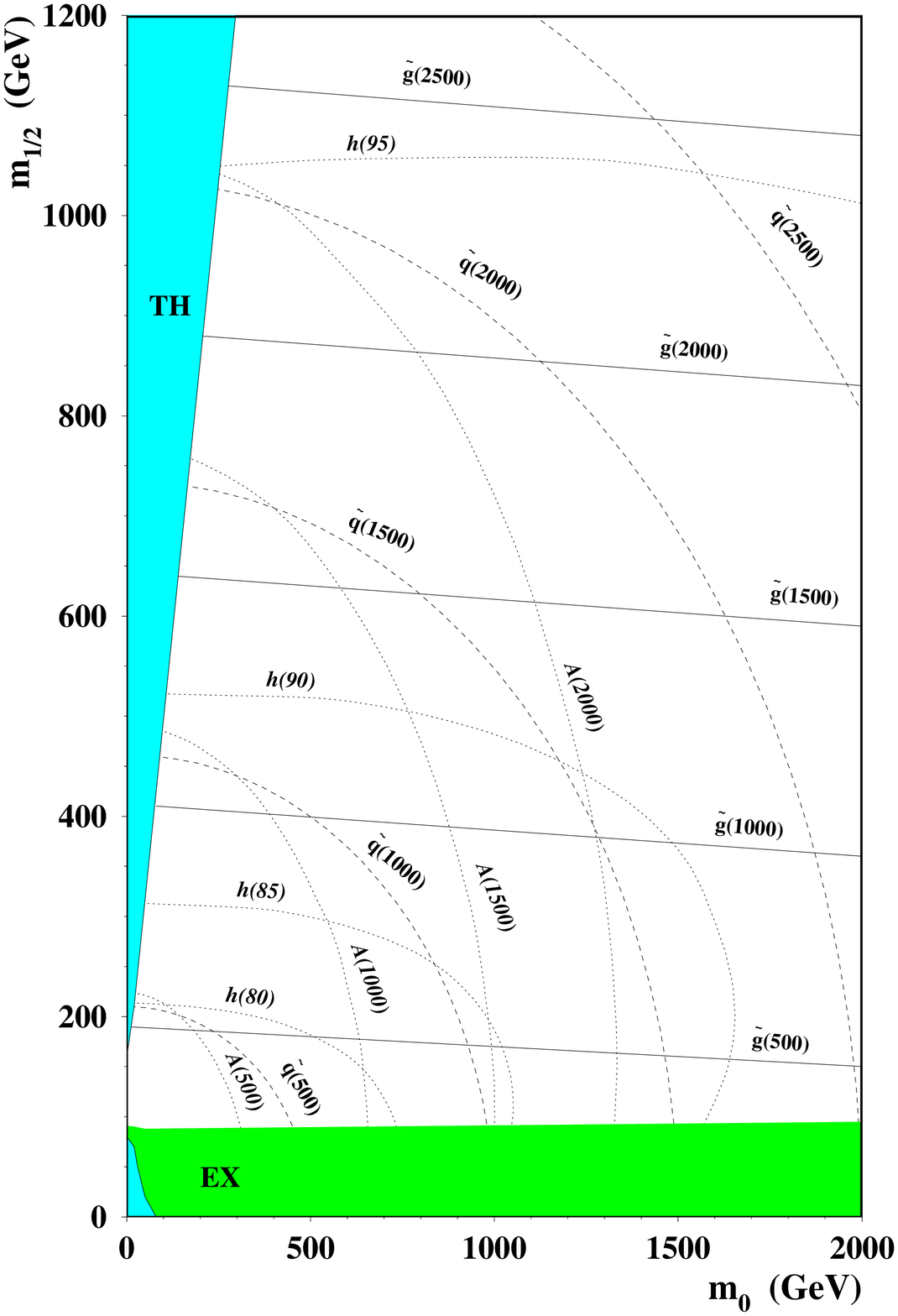}} & ~ &
\vspace*{0mm}
\hspace*{-5mm}\resizebox{.465\textwidth}{10cm}
                      {\includegraphics{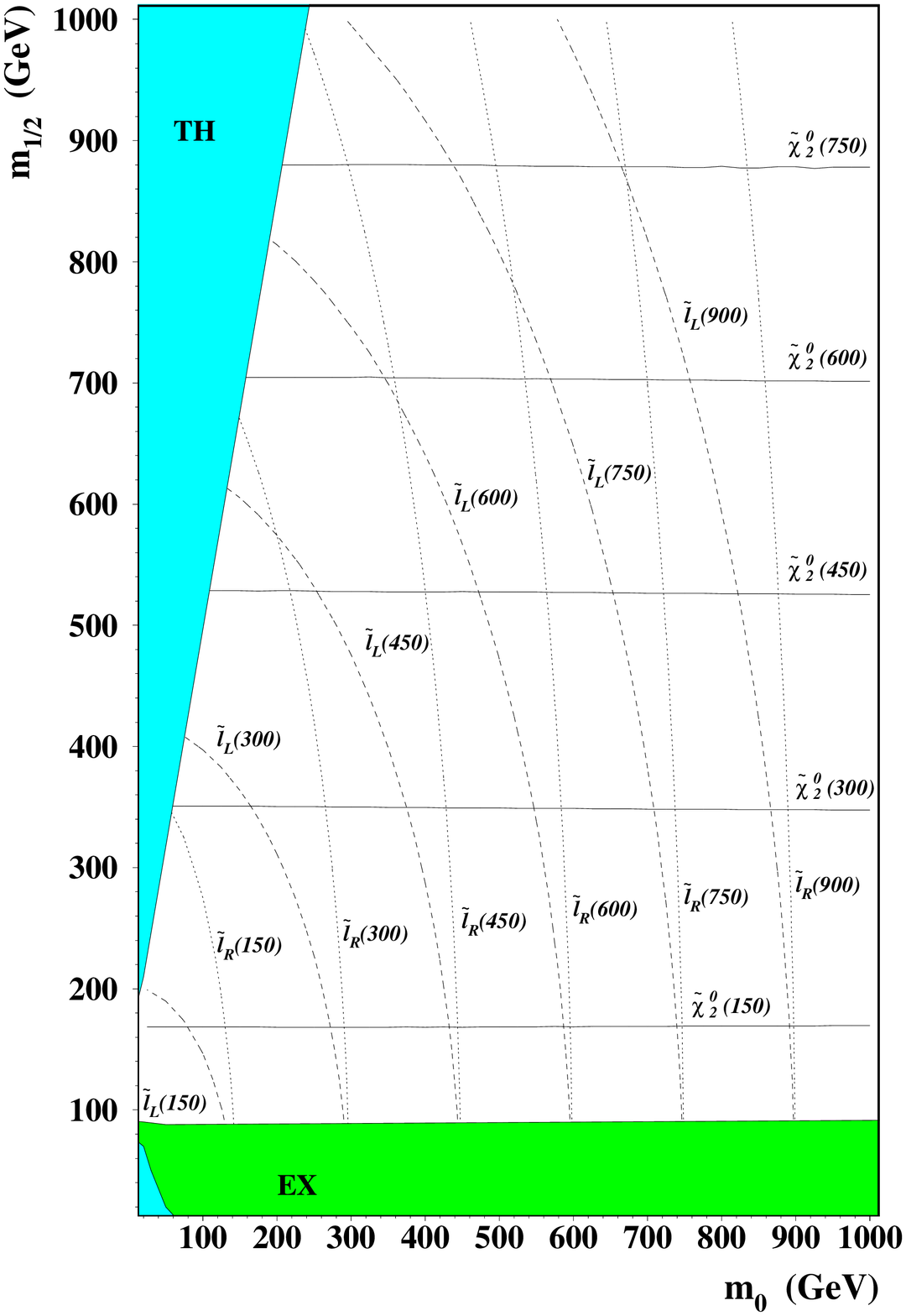}} \\
\vspace*{-5mm}
 \caption{ Isomass contours of squarks($\tilde{q}$), gluino($\tilde{g}$),
 lightest scalar(h) and 
     pseudoscalar(A) Higgs fields as a function of
     mSUGRA parameters m$_0$,  m$_{1/2}$ for fixed values of 
    tan$\beta$=2, $\mu$$<$0 and A$_0$=0. Numbers in parenthesis denote 
    the masses in GeV.}
    &~&
\vspace*{-5mm}
 \caption{Isomass contours of left($\tilde{l}_{L}$)
          and right($\tilde{l}_{R}$) sleptons and next-to-lightest
          neutralino($\tilde{\chi}_{2}^{0}$)
 with the same choise of mSUGRA parameters as in Fig.1.
 Note expanded scale compared to Fig.1. }
\end{myfig2}
\begin{table}[htb]
  \caption{Sets of mSUGRA parameter values investigated.}
  \label{tab:1}
  \begin{center}
    \renewcommand{\arraystretch}{1.2}
    \setlength{\tabcolsep}{2mm}
\begin{tabular}{|c|c|c|} \hline
 mSUGRA parameter &             &              \\
    Set       & tan$\beta$  & sign($\mu$)  \\
\hline \hline
   1 &  2 & -1 \\
\hline
   2 &  2 & +1 \\
\hline
   3 & 35 & -1 \\ 
\hline
   4 & 35 & +1 \\
\hline
\end{tabular}
\end{center}
\end{table}
\vspace*{3mm}
\begin{myfig2}{hbtp}
\vspace*{-5mm}
\hspace*{-5mm}\resizebox{.465\textwidth}{9cm}
                        {\includegraphics{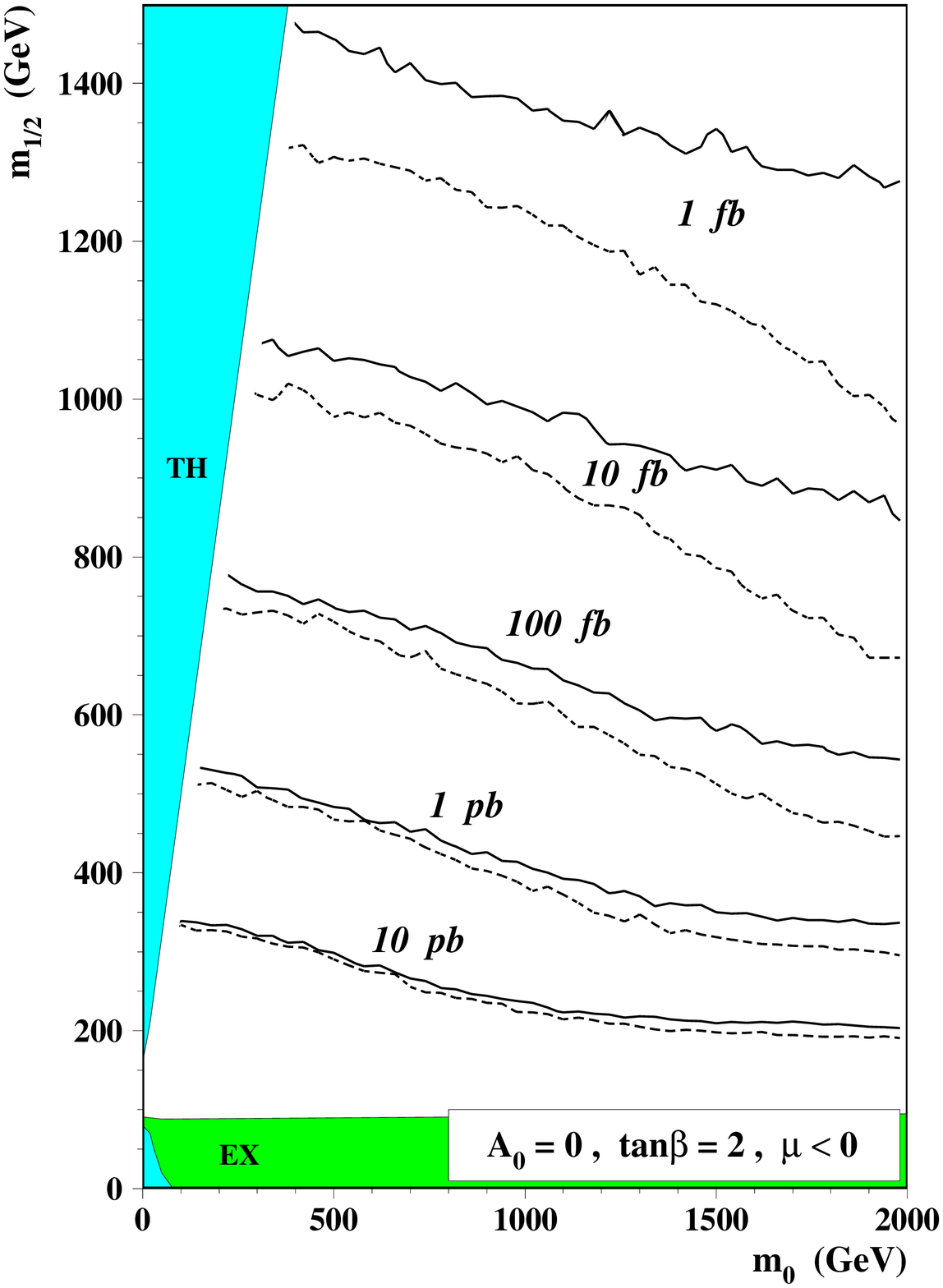}} & ~ &
\vspace*{-5mm}
\hspace*{-5mm}\resizebox{.465\textwidth}{9cm}
                        {\includegraphics{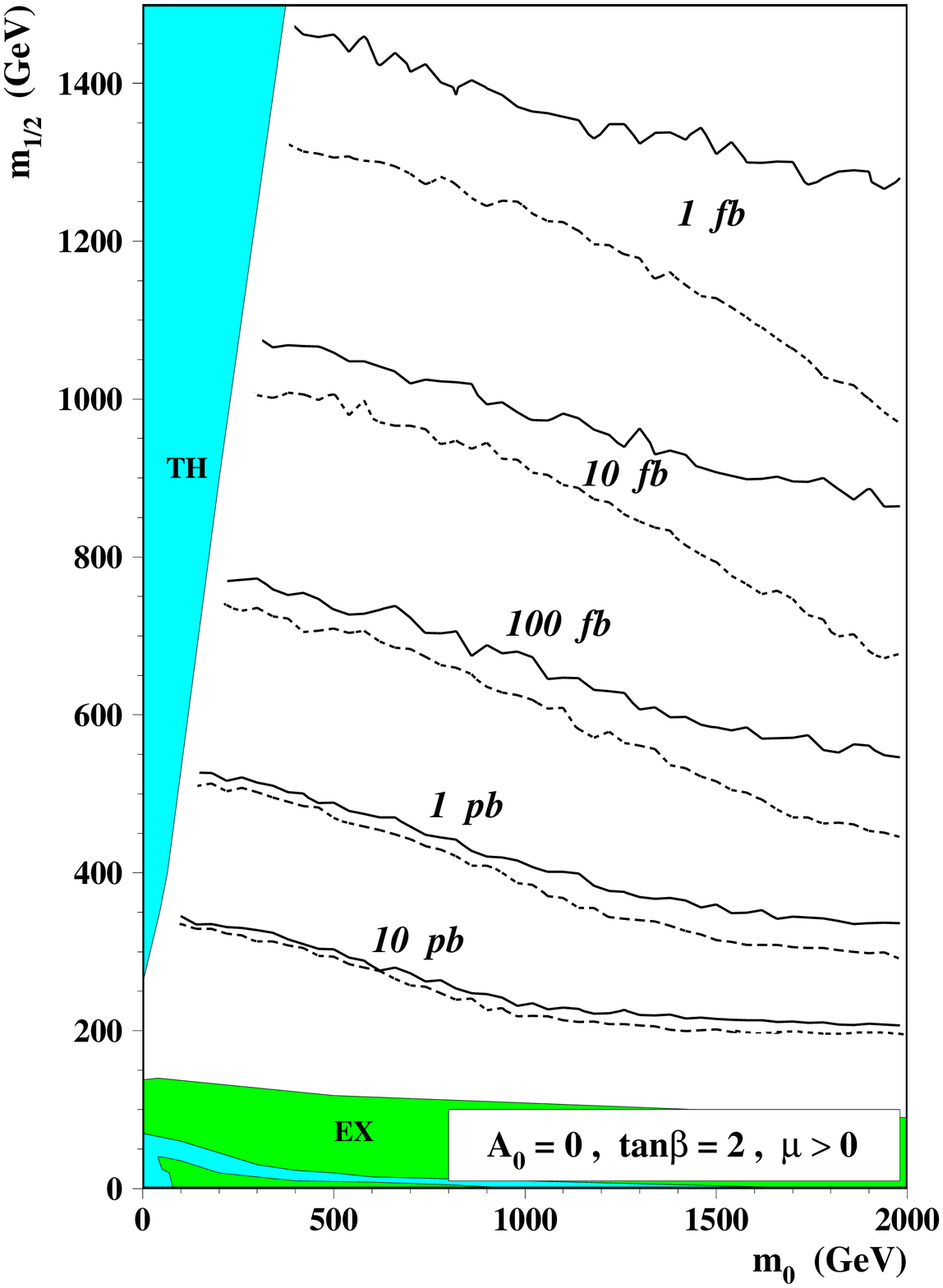}} \\
\vspace*{-5mm}
 \caption{ Total mSUGRA cross-section contours as a function of         
                       m$_0$, m$_{1/2}$  for mSUGRA Set 1 (solid line).
    The contribution
    of $\tilde{g}\tilde{g}$, $\tilde{g}\tilde{q}$, $\tilde{q}\tilde{q}$
    production alone is shown by dashed line.}    &~&
\vspace*{-5mm}
 \caption{ Same as Fig.3, except for Set 2.}
\end{myfig2}
\vspace*{-10mm}
\begin{myfig2}{hbtp}
\vspace*{-5mm}
\hspace*{-5mm}\resizebox{.465\textwidth}{9cm}
                        {\includegraphics{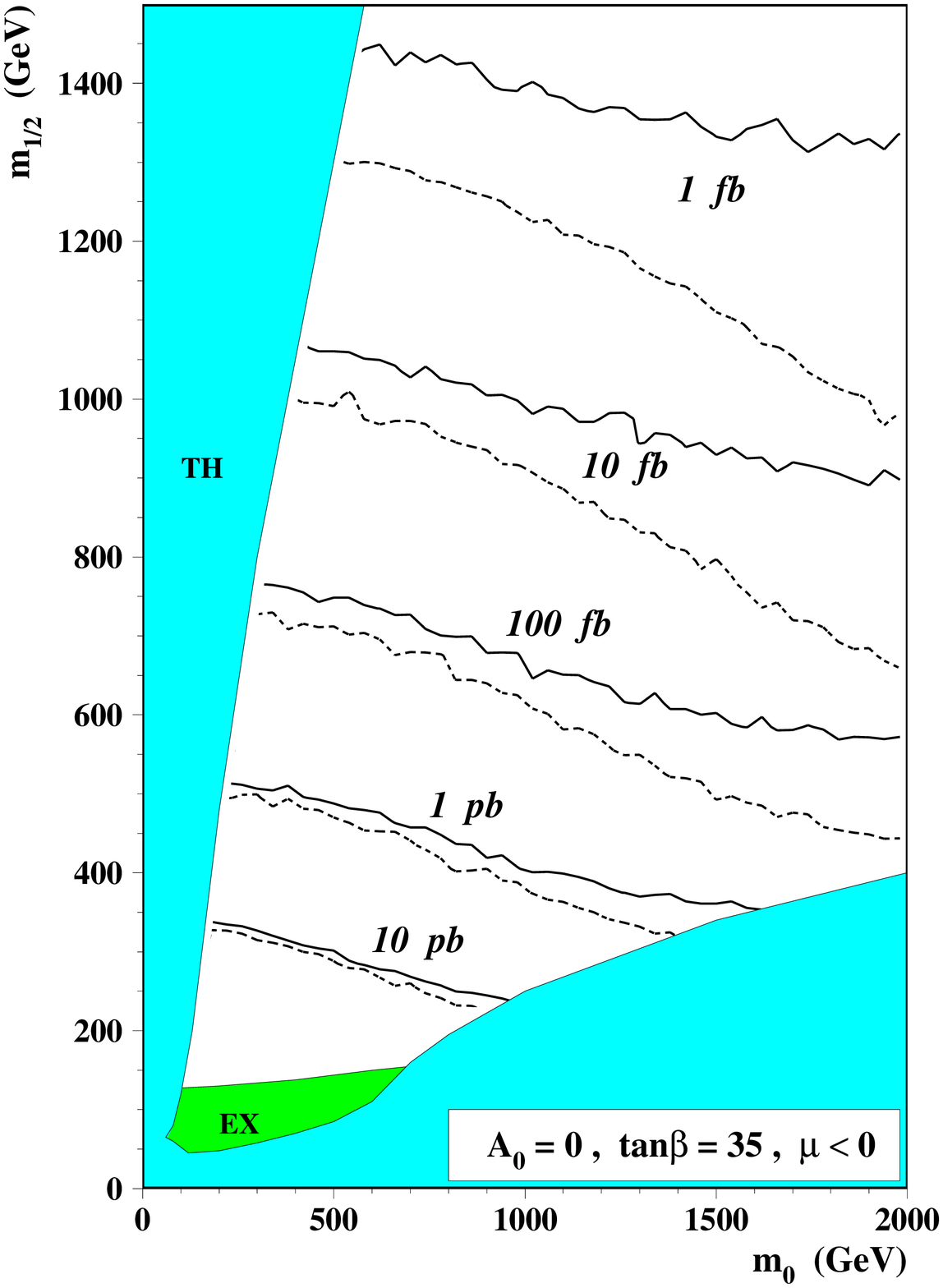}} & ~ &
\vspace*{-5mm}
\hspace*{-5mm}\resizebox{.465\textwidth}{9cm}
                        {\includegraphics{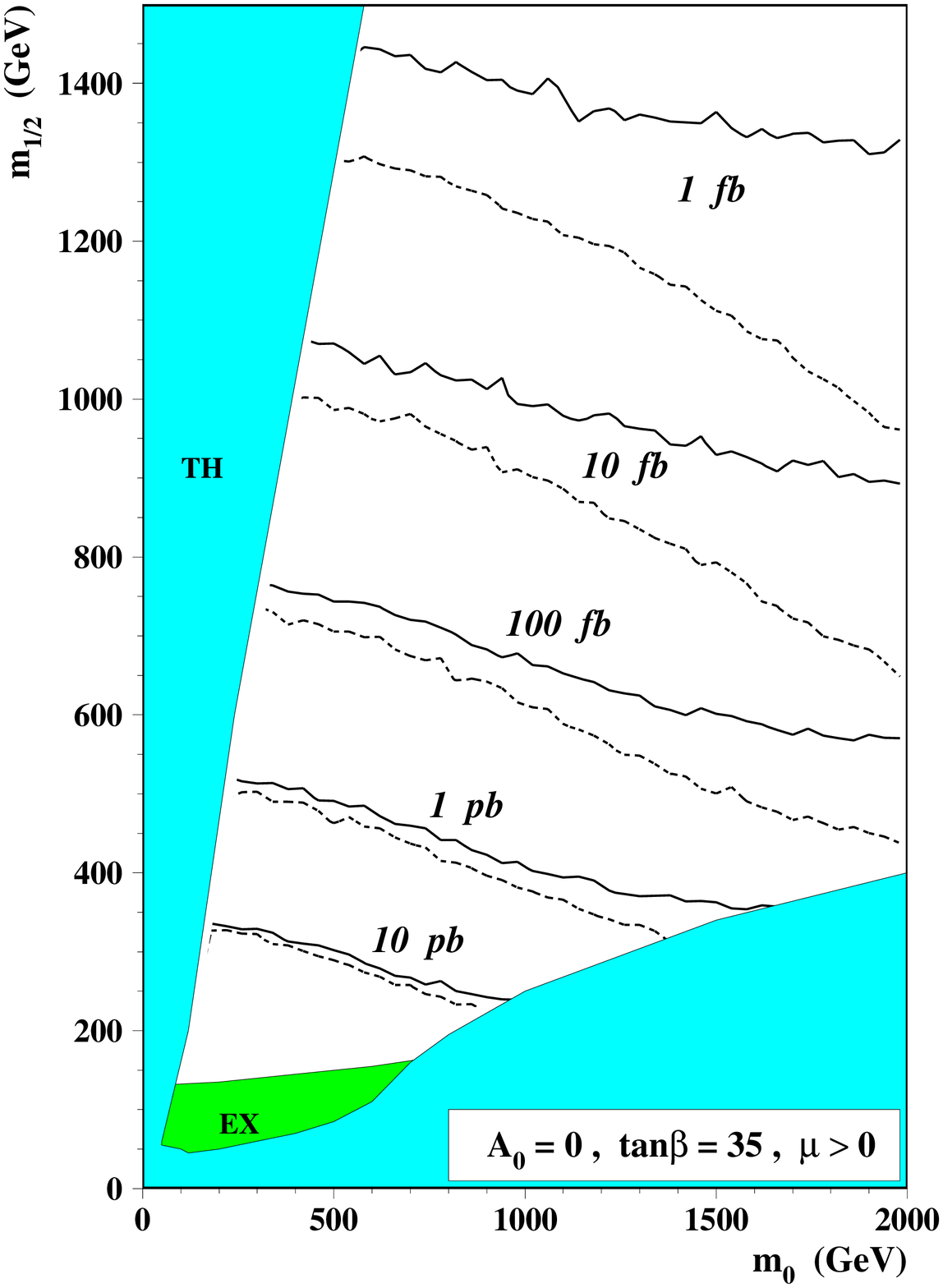}} \\
\vspace*{-5mm}
 \caption{ Same as Fig.3, except for Set 3.}
    &~&
\vspace*{-5mm}
 \caption{ Same as Fig.3, except for Set 4.}
\end{myfig2}

In Figs.3-6
 one can see total mSUGRA production cross-section as a function of 
m$_0$, m$_{1/2}$ for chosen sets of  tan$\beta$ and sign($\mu$).
The contribution of strongly interacting SUSY particles cross-section
is also shown by dashed line.  
The jitter of contours is caused by limited statistics. The total 
cross-section  for different values of tan$\beta$ and sign($\mu$), but
for the same values of m$_0$, m$_{1/2}$ differs slightly.
The bulk of the total cross-section for low values of m$_{1/2}$ consists from
$\tilde{g}\tilde{g}$, $\tilde{g}\tilde{q}$, $\tilde{q}\tilde{q}$, whereas
in the domains with extremely high masses of $\tilde{g}$, $\tilde{q}$  
the contribution of production of squarks or gluinos  associated
with charginos and neutralinos may dominate.

Figs.7 and 8
 shows the typical decay modes of heavy gluino and left squark in 
case of high tan$\beta$, when $\tilde{\chi}_{2}^{0}$ and
  $\tilde{\chi}_{1}^{\pm}$  branching ratios for 
decays into $\tilde{\tau}_{1,2}$ $\tau(\nu)$  exceed 60 $\%$ due to 
large tau Yukava couplings \cite{large_tan}.
To simplify the figure, similar intermediate final states were joined. For
instance, states 
 $\tilde{\chi}_{2}^{\pm}$Wbb and  $\tilde{\chi}_{2}^{\pm}$tb
were treated(summed up) as the same, though they have
different kinematics, in principle (see the rightmost round mark at
 $\tilde{\chi}_{2}^{\pm}$ horizontal line (1067 GeV).
 It is almost impossible to follow and 
calculate all the branchings for gluino decays,
 so some small ones are not shown thus resulting
in some small underestimate of the ``final'' states (at the level of  
 $\tilde{\chi}_{1}^{0}$) branching ratios.
The final states having the highest branching ratios 
are listed in the lower part of 
Figs.7 and 8.
The right squarks ($\tilde{q}_R$) decay entirely
into  $\tilde{\chi}_{1}^{0}$ q  final state in the domain of mSUGRA 
parameter space where m$_{\tilde{q}_{R}}<$m$_{\tilde{g}}$ as it is in
the point presented in
 Figs.7 and 8. 
Decay chains of $\tilde{b}_{1,2}$ or $\tilde{t}_{1,2}$
 are somewhat intermediate 
between those for $\tilde{q}_{L}$ and  $\tilde{g}$ from the point of view of
variety of final states.

Figs.9 and 10
 show typical decay modes of a heavy gluino and left squark
 respectively, at the same point of parameter space as in 
Figs.7 and 8,
 except for low  tan$\beta$=2. Right squarks again decay entirely
into LSP + quarks.
One can see that decay chains of gluino are not so complicated in case of low 
tan$\beta$, mainly due to the fact that m$_{\tilde{t}_1}<$m$_{
\tilde{\chi}_{2}^{\pm},\tilde{\chi}_{3,4}^{0}}$. In addition,
 at low tan$\beta$ 
$\tilde{\tau}_{1,2}$ do not dominate in the decays of  
  $\tilde{\chi}_{1}^{\pm}$ and $\tilde{\chi}_{2}^{0}$, instead,
branchings of  $\tilde{\chi}_{1}^{\pm}$ and $\tilde{\chi}_{2}^{0}$ decays into
sleptons are enhanced. So final states of left squarks and gluino contain
more leptons in case of low  tan$\beta$ than in case of high  tan$\beta$
 in the chosen 
particular point of mSUGRA parameter space. 
The  latter statement is more general, namely this difference in the yield of
leptons between low and high tan$\beta$ exists in significant domains of
 m$_0$, m$_{1/2}$ values along the theoretically excluded region at low
values of m$_0$, where $\tilde{\chi}_{1}^{\pm}$ and $\tilde{\chi}_{2}^{0}$  
have 2-body decays.
\begin{figure}[hbtp]
\begin{center}
\vspace*{0mm}
\hspace*{0mm}\resizebox{0.99\textwidth}{20cm}
                        {\includegraphics{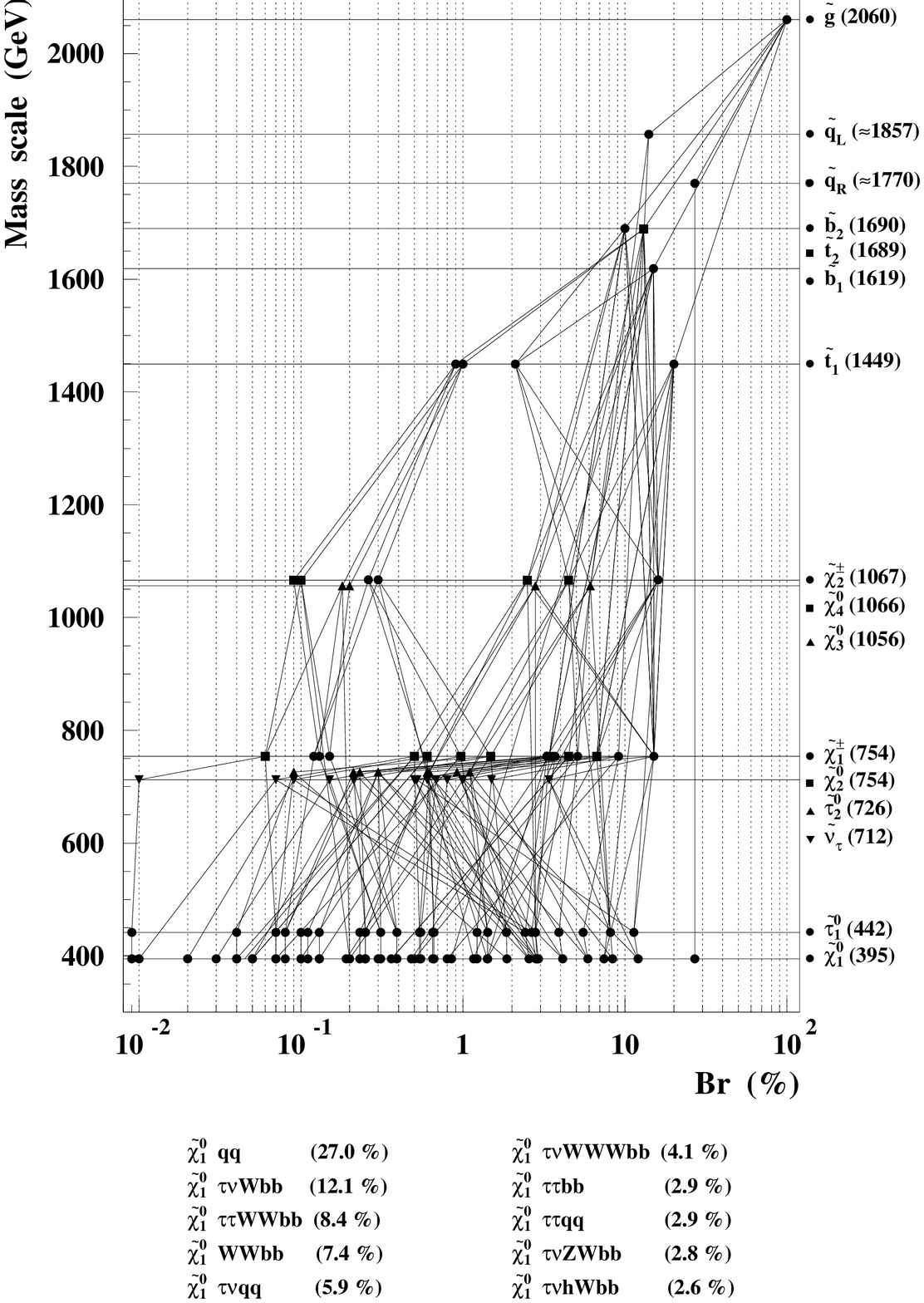}}
 \caption{ Typical decay modes for massive (2060 GeV) gluino for high 
           tan$\beta$ ( m$_0$=400 GeV, m$_{1/2}$=900 GeV, Set 4).}
\end{center}
\end{figure}
\begin{figure}[hbtp]
\begin{center}
\vspace*{0mm}
\hspace*{0mm}\resizebox{0.99\textwidth}{20cm}
                        {\includegraphics{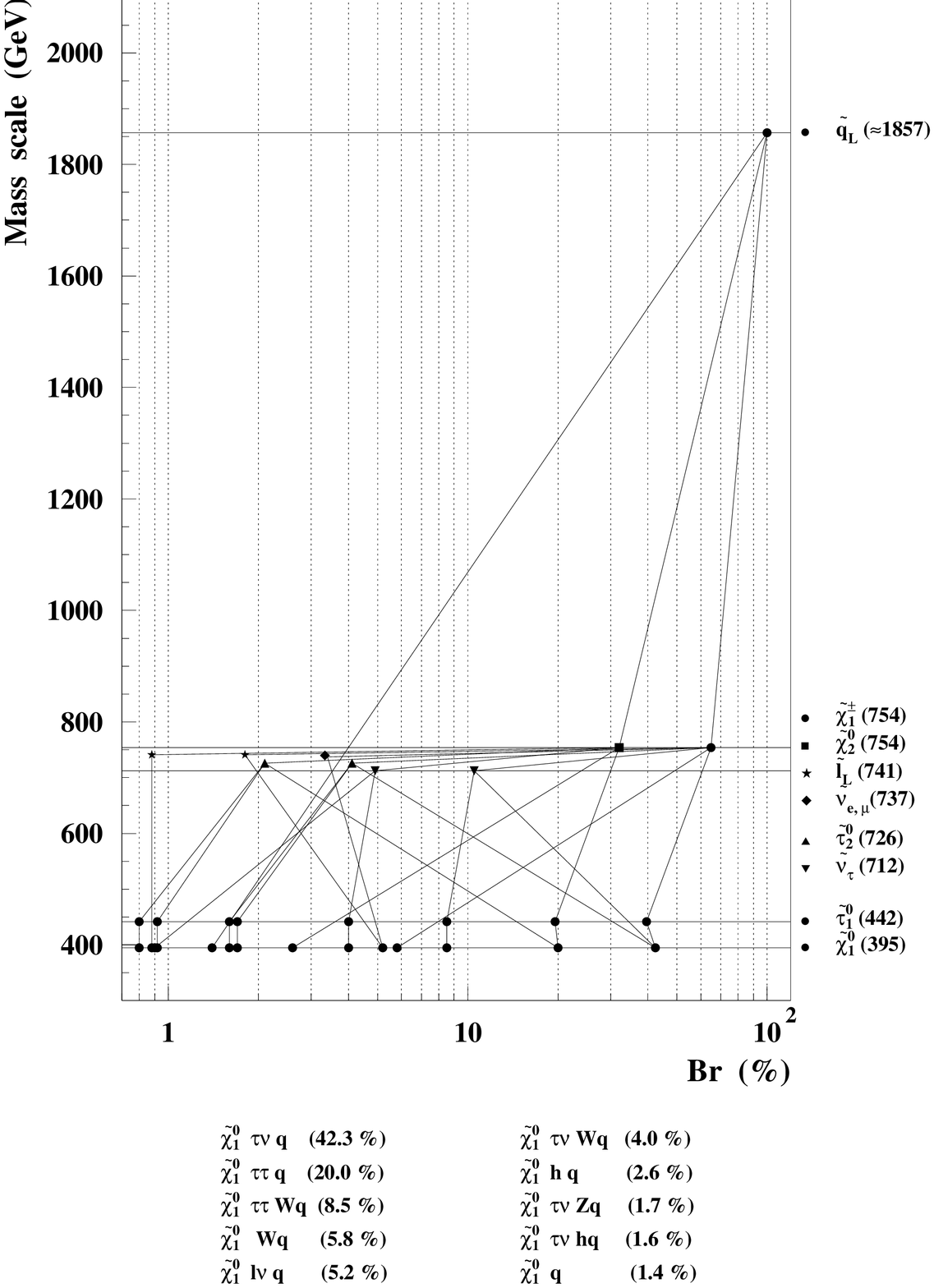}}
 \caption{ Same as Fig.7, but for first generation left squarks (1857 GeV)}
\end{center}
\end{figure}
\begin{figure}[hbtp]
\begin{center}
\vspace*{0mm}
\hspace*{0mm}\resizebox{0.99\textwidth}{20cm}
                        {\includegraphics{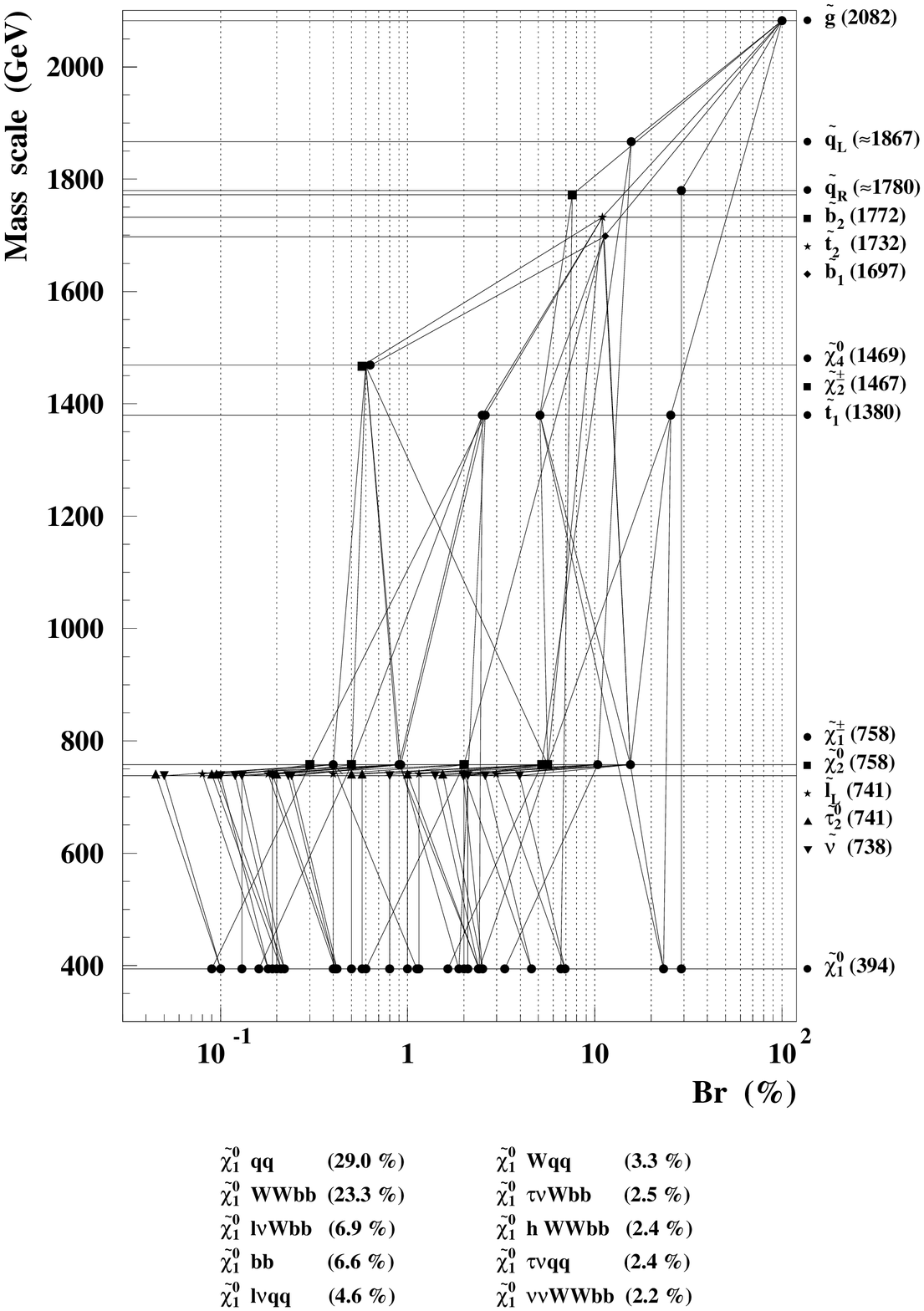}}
 \caption{ Typical decay modes for massive (2082 GeV) gluino for low  
           tan$\beta$ ( m$_0$=400 GeV, m$_{1/2}$=900 GeV, Set 2).}
\end{center}
\end{figure}
\begin{figure}[hbtp]
\begin{center}
\vspace*{0mm}
\hspace*{0mm}\resizebox{0.99\textwidth}{20cm}
                        {\includegraphics{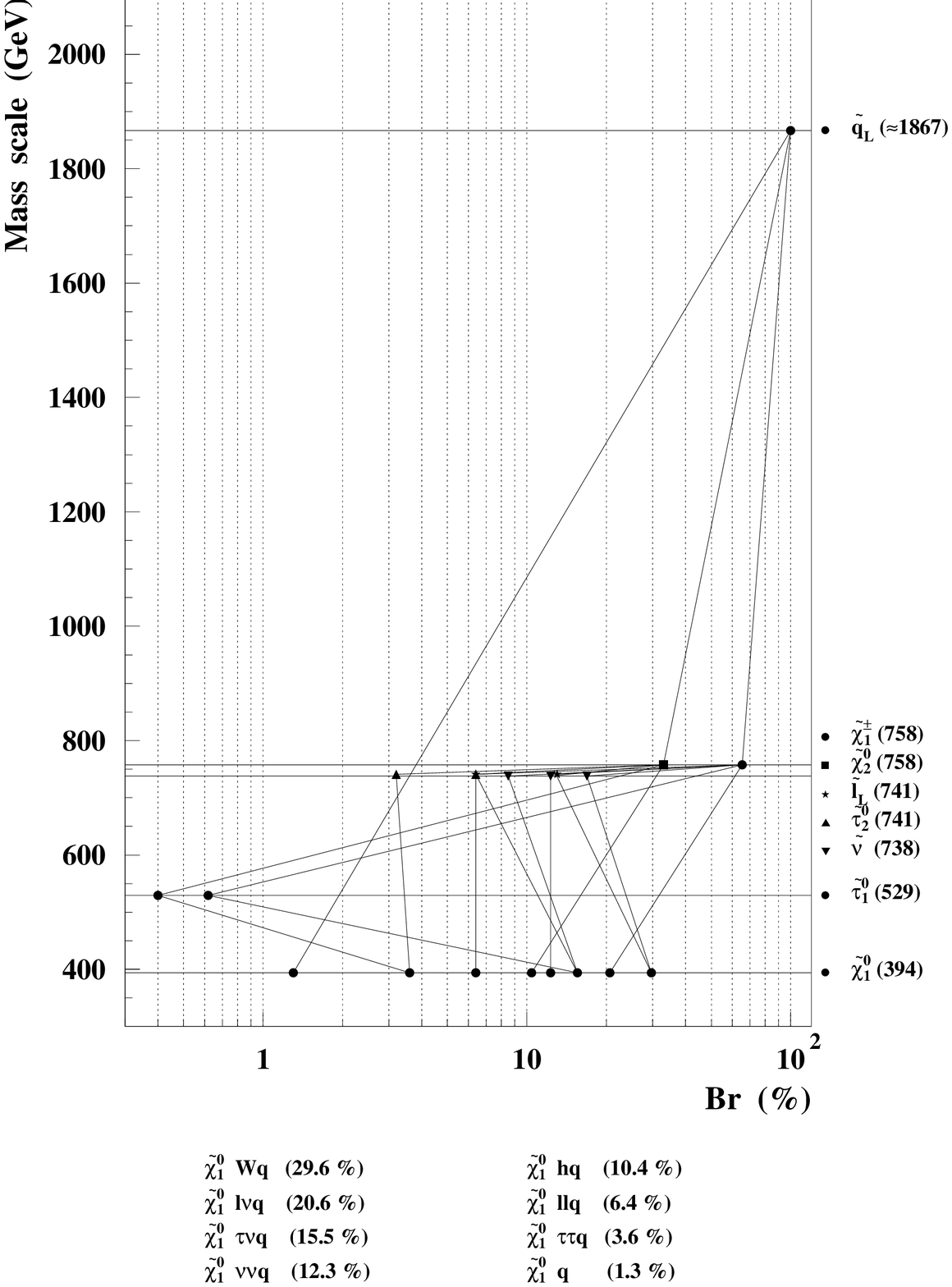}}
 \caption{ Same as Fig.9, but for first generation left squarks (1867 GeV)}
\end{center}
\end{figure}
\begin{myfig2}{hbtp}
\vspace*{-10mm}
\hspace*{-5mm}\resizebox{.465\textwidth}{9.5cm}
                        {\includegraphics{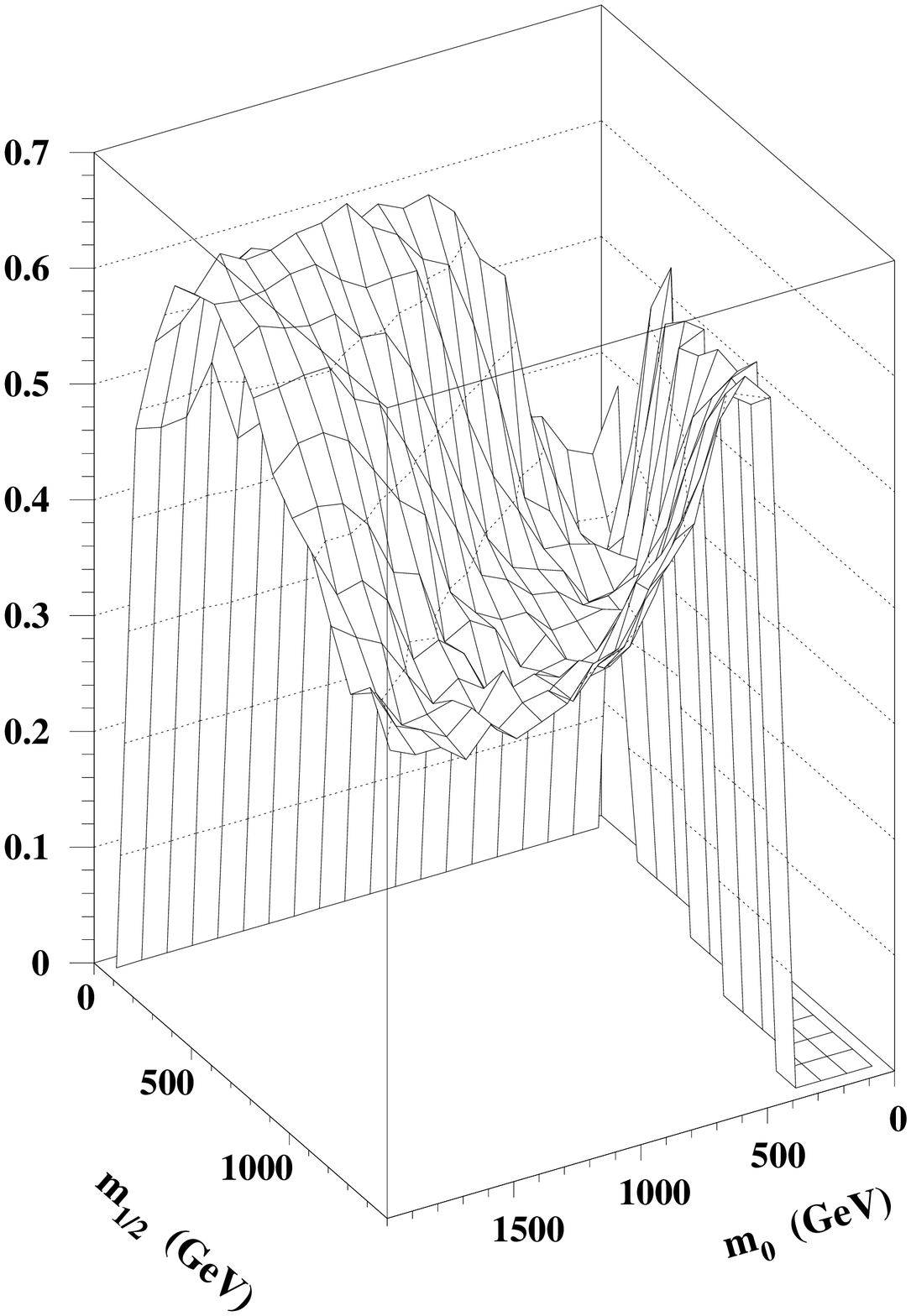}} & ~ &
\vspace*{-10mm}
\hspace*{-5mm}\resizebox{.465\textwidth}{9.5cm}
                        {\includegraphics{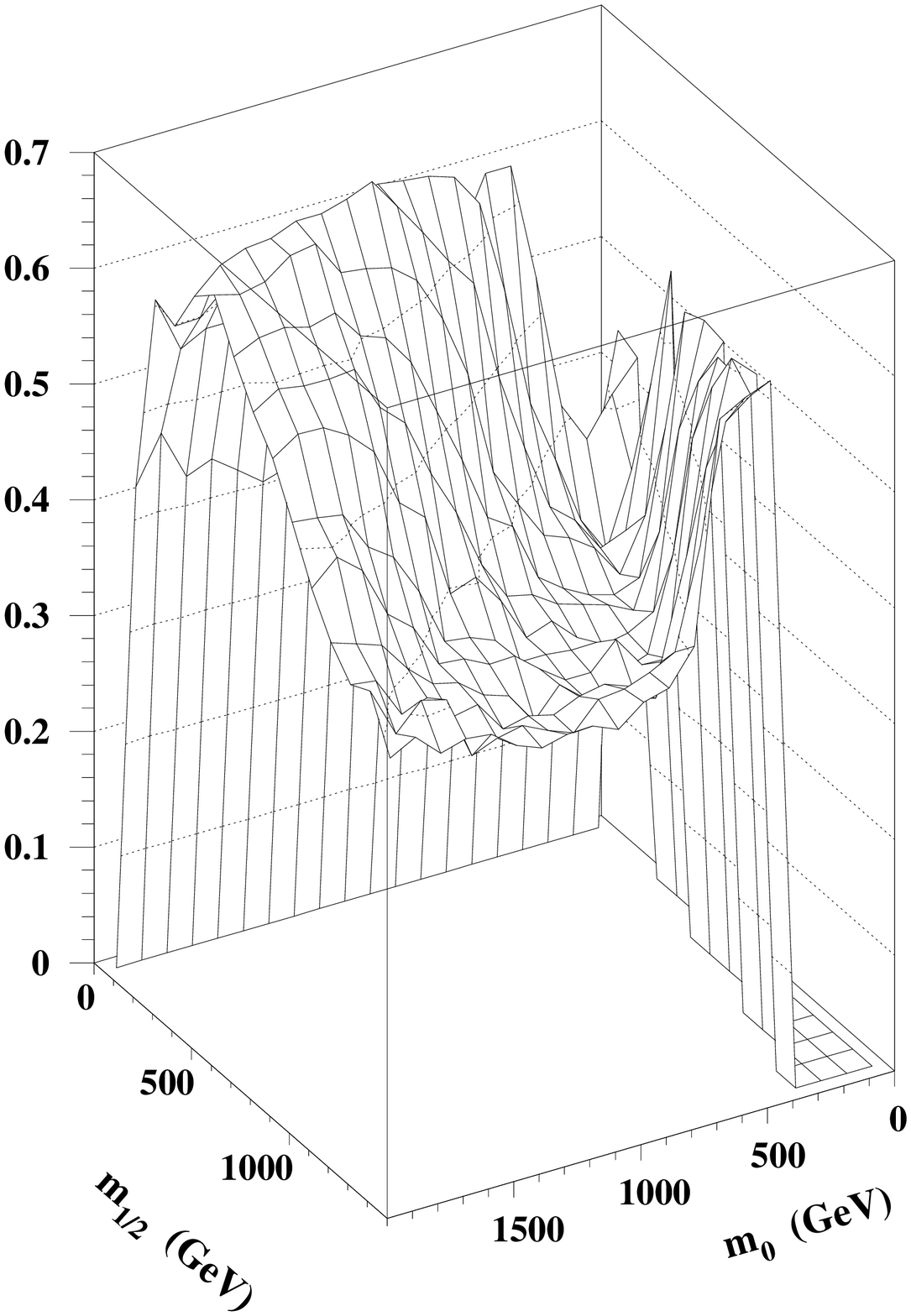}} \\
\vspace*{-5mm}
 \caption{ Probability to find at least one lepton in the final state 
           (e or $\mu$ as defined in text) 
                 in signal evens as a function of
           m$_0$, m$_{1/2}$  for mSUGRA Set 1.}
    &~&
\vspace*{-5mm}
\caption{ Same as Fig.11, except for Set 2.} 
\end{myfig2}
\begin{myfig2}{hbtp}
\vspace*{-10mm}
\hspace*{-5mm}\resizebox{.465\textwidth}{9.5cm}
                        {\includegraphics{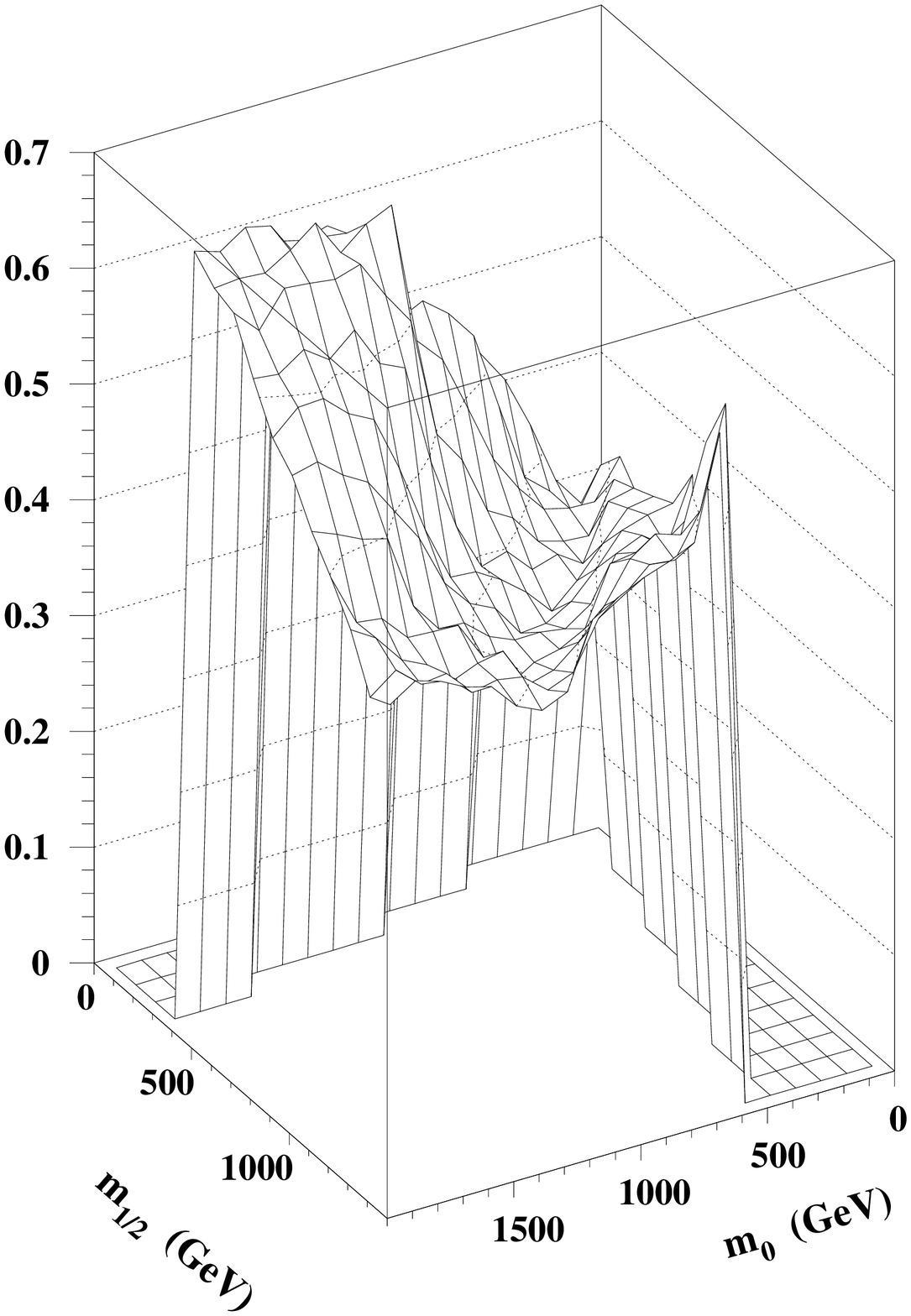}} & ~ &
\vspace*{-10mm}
\hspace*{-5mm}\resizebox{.465\textwidth}{9.5cm}
                        {\includegraphics{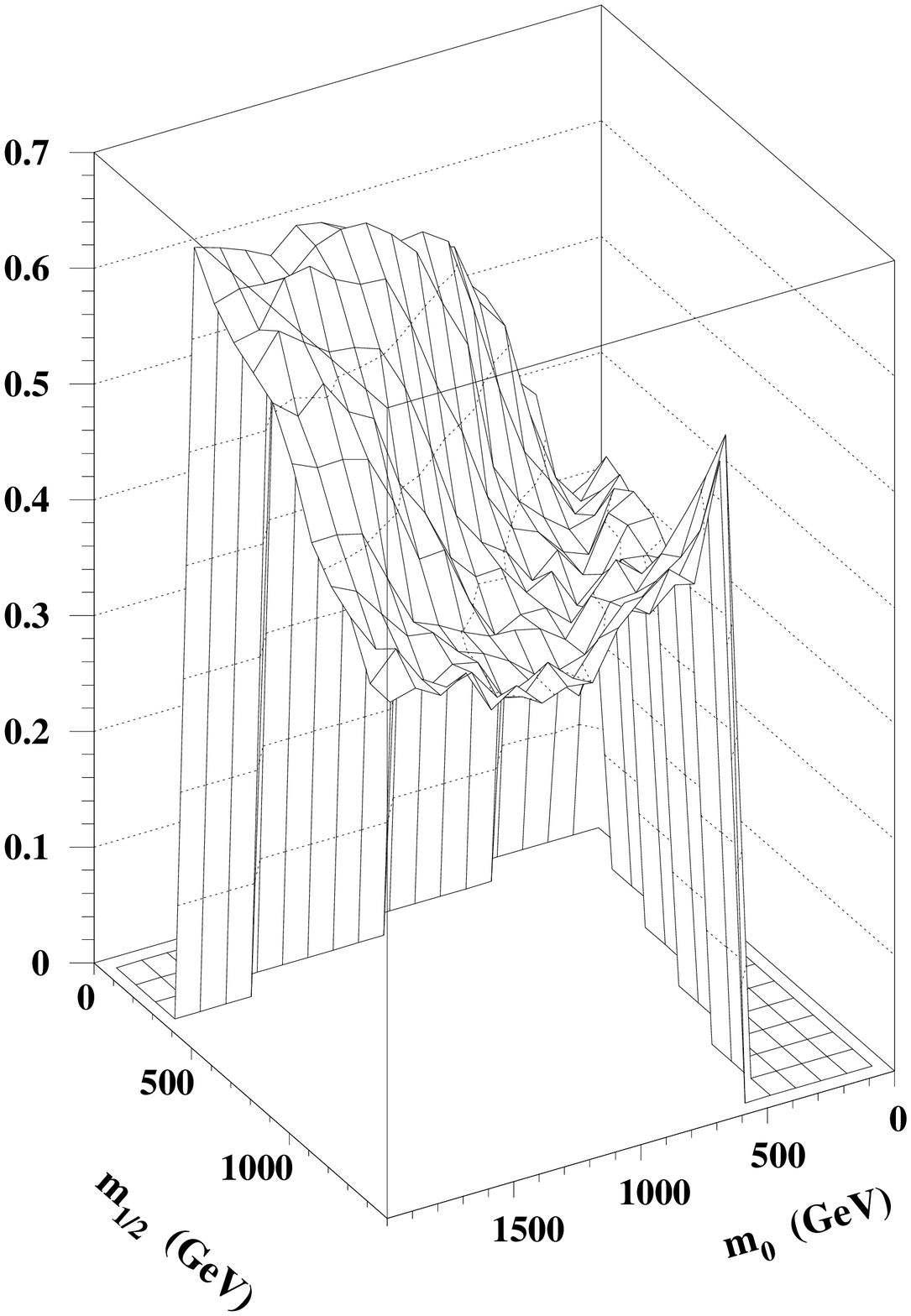}} \\
\vspace*{-5mm}
 \caption{ Same as Fig.11, except for Set 3.}
    &~&
\vspace*{-5mm}
\caption{ Same as Fig.11, except for Set 4.}
\end{myfig2}

Figs.11-14
 illustrate the fact that mSUGRA final states frequently contain
lepton(s). The plots show the probability to find at least one lepton
per mSUGRA event above
some p$_{T}$ threshold (10 GeV for muons, 20 GeV for electrons)
 within the detector acceptance ( $|\eta| <$ 2.4).
 The source of these leptons
are mainly b-jets produced in the decay chains of sparticles, then
W and Z-bosons produced both in decays of top and chargino/neutralino 
decays. 
One of the abundant sources of leptons in mSUGRA final states are also
 leptonic decays (2-body via sleptons or direct 3-body)
of charginos and neutralinos (mainly $\tilde{\chi}_{2}^{0}$, 
 $\tilde{\chi}_{1}^{\pm}$) in some domains of the parameter space.
One can see the domain where $\tilde{\chi}_{2}^{0}$, $\tilde{\chi}_{1}^{\pm}$ 
have significant branching ratio for the decays into sleptons $\rightarrow$
 leptons on the right side of
Figs.11 and 12
 (small m$_{0}$ values). A similar situation,
but not so pronounced can be seen in
 Figs.13 and 14,  where the spike in the vicinity of m$_{0}$=600 GeV, 
m$_{1/2}$=1500 GeV reflects the increased (with increase of m$_{1/2}$)
branchings of 
 $\tilde{\chi}_{2}^{0}$,  $\tilde{\chi}_{1}^{\pm}$  into $\mu$,e-sleptons,
 thus replacing high branchings of 
 $\tilde{\chi}_{2}^{0}$,  $\tilde{\chi}_{1}^{\pm}$
into $\tilde{\tau}_{1,2}$ at low values of  m$_{0}$, m$_{1/2}$.
 In the mentioned extreme point with 
 m$_{0}$=600 GeV, m$_{1/2}$=1500 GeV  we have 
Br($\tilde{\chi}_{2}^{0}$ $\rightarrow$
$\tilde{\chi}_{1}^{0}$ $l^{+}l^{-}$) = 23 \% and
 Br($\tilde{\chi}_{1}^{\pm}$ $\rightarrow$ $\tilde{\chi}_{1}^{0}$ $l^{\pm}\nu$)
= 42 \%.  

All the figures showed in this section are drawn using calculations made with
ISAJET 7.32 generator \cite{ISAJET} and supplements therein.
One can also take a look at the relevant figures of mSUGRA events
 in CMS detector
selected and reconstructed with fast MC code called CMSJET \cite{CMSJET}
used in this study (see also section 3)
and then drawn with CMSIM \cite{CMSIM} GEANT-based CMS detector simulation
package.
Figs.2 and 10 in \cite{pictures} are  $\tilde{g}$, $\tilde{q}$ events with
different final states in two distant points in mSUGRA parameter space.
We do not show them here because of the extremely large size of these
drawings (some 12 Mb).

\section{Simulation procedure}

The PYTHIA 5.7 generator \cite{PYTHIA} is used to generate  all
 SM background processes, whereas ISAJET 7.32 
is used for mSUGRA signal simulations.
    The CMSJET (version 4.51) fast MC package \cite{CMSJET}
 is used to model the CMS detector \cite{CMS} response, since it still looks
impossible to perform a full-GEANT simulation for the present study,
 requiring to 
process multi-million samples of signal and SM background events.
A sketch of the calorimeter model implemented in CMSJET is shown in 
Fig.15. The sign ``+'' in the $\sigma$/E expressions
 means sum in quadrature everywhere it appears in this figure.

The SM background processes considered are $:$  QCD  2 $\rightarrow$ 2
 (including $b\bar{b}$), $t\bar{t}$, $W+jets$, $Z+jets$. The $\hat{p}_{T}$
range of all the background processes is subdivided into several  intervals to
facilitate  accumulation of statistics in the high-$\hat{p}_{T}$ range $:$
100-200 GeV, 200-400 GeV, 400-800 GeV and  $>$ 800 GeV (additional interval
of 800-1200 GeV is reserved for QCD).
The accumulated SM background statistics for all background channels
is presented in Tab.2, whilst the signal data samples are given in Tab.3.

The grid of probed m$_{0}$, m$_{1/2}$ mSUGRA points has a cell size of
 $\Delta$m$_{0}$=$\Delta$m$_{1/2}$=100 GeV for m$_{0}<$1000 GeV and 
 $\Delta$m$_{0}$=200 GeV,
 $\Delta$m$_{1/2}$=100 GeV for  m$_{0}>$1000 GeV. 
Set 4 was also probed with the appropriate mixture of signal and pile-up 
events (see details in subsection 6.4). 
\begin{table}[htb]
  \caption{SM background statistics generated.}
  \label{tab:2}
  \begin{center}
    \renewcommand{\arraystretch}{1.2}
    \setlength{\tabcolsep}{2mm}
\begin{tabular}{|c|c|c|c|c|} \hline
  Bkgd channel & $\hat{P}_{T}$ interval & $\sigma$ (pb) & N$_{ev}$ generated &
  \% of needed   \\
              & (GeV) & (pb) & & for 100 fb$^{-1}$ \\
 \hline \hline
            &   0 - 100 & 267       &  1.461$\cdot$10$^{7}$  &  54.7 \\ 
            & 100 - 200 & 240       &  6.638$\cdot$10$^{6}$  &  27.7 \\
 $t\bar{t}$ & 200 - 400 & 80.7      &  6.864$\cdot$10$^{6}$  &  85.1 \\
            & 400 - 800 &  6.3      &  6.484$\cdot$10$^{5}$  & 102.9 \\
            & $>$ 800   & 0.163     &  1.630$\cdot$10$^{4}$  & 100.0 \\
\hline
            &  50 - 100 & 2670      &  1.554$\cdot$10$^{7}$  &   5.8 \\ 
            & 100 - 200 & 580       &  9.998$\cdot$10$^{6}$  &  17.2 \\
  $Zj$      & 200 - 400 & 64.0      &  4.455$\cdot$10$^{6}$  &  71.2 \\
            & 400 - 800 &  4.0      &  4.927$\cdot$10$^{5}$  & 123.2 \\
            & $>$ 800   & 0.137     &  1.370$\cdot$10$^{4}$  & 100.0 \\
\hline
            &  50 - 100 & 7140      &  2.753$\cdot$10$^{7}$  &   3.9 \\ 
            & 100 - 200 & 1470      &  8.618$\cdot$10$^{6}$  &   5.9 \\
  $Wj$      & 200 - 400 & 155       &  6.424$\cdot$10$^{6}$  &  41.4 \\
            & 400 - 800 &  9.5      &  9.909$\cdot$10$^{5}$  & 104.3 \\
            & $>$ 800   & 0.33      &  3.300$\cdot$10$^{4}$  & 100.0 \\
\hline
            & 100 - 200  & 1.37$\cdot$10$^6$ & 6.000$\cdot$10$^{7}$  &   0.04\\
            & 200 - 400  & 7.15$\cdot$10$^4$ & 3.229$\cdot$10$^{7}$  &   0.45\\
 $QCD$      & 400 - 800  & 2740              & 3.259$\cdot$10$^{7}$  &  11.9 \\
(incl. $b\bar{b}$)& 800 - 1200 & 60.0        & 6.033$\cdot$10$^{6}$  & 100.5 \\
            & $>$ 1200   &  4.8              & 4.947$\cdot$10$^{5}$  & 103.1 \\
\hline
\hline
 total      &            &                   & 2.342$\cdot$10$^{8}$  &    \\
\hline
\end{tabular}
\end{center}
\end{table}
\begin{table}[htb]
  \caption{mSUGRA signal statistics generated.}
  \label{tab:3}
  \begin{center}
    \renewcommand{\arraystretch}{1.2}
    \setlength{\tabcolsep}{2mm}
\begin{tabular}{|c|c|c|} \hline
  Set No. &  No. of probed  m$_{0}$, m$_{1/2}$ points & Total statistics 
generated \\
 \hline \hline
      1     &       120             &    1.95$\cdot$10$^{6}$         \\
\hline
      2     &       114             &    1.87$\cdot$10$^{6}$         \\
\hline
      3     &       99              &    0.68$\cdot$10$^{6}$         \\
\hline
      4     &       100             &    0.75$\cdot$10$^{6}$         \\
\hline
4 with pile-up &    67              &    0.18$\cdot$10$^{5}$         \\
\hline
\hline
 total      &      500              &    5.43$\cdot$10$^{6}$         \\
\hline
\end{tabular}
\end{center}
\end{table}
\begin{figure}[hbtp]
\begin{center}
\vspace*{-10mm}
\hspace*{0mm}\resizebox{0.99\textwidth}{21.5cm}
                        {\includegraphics{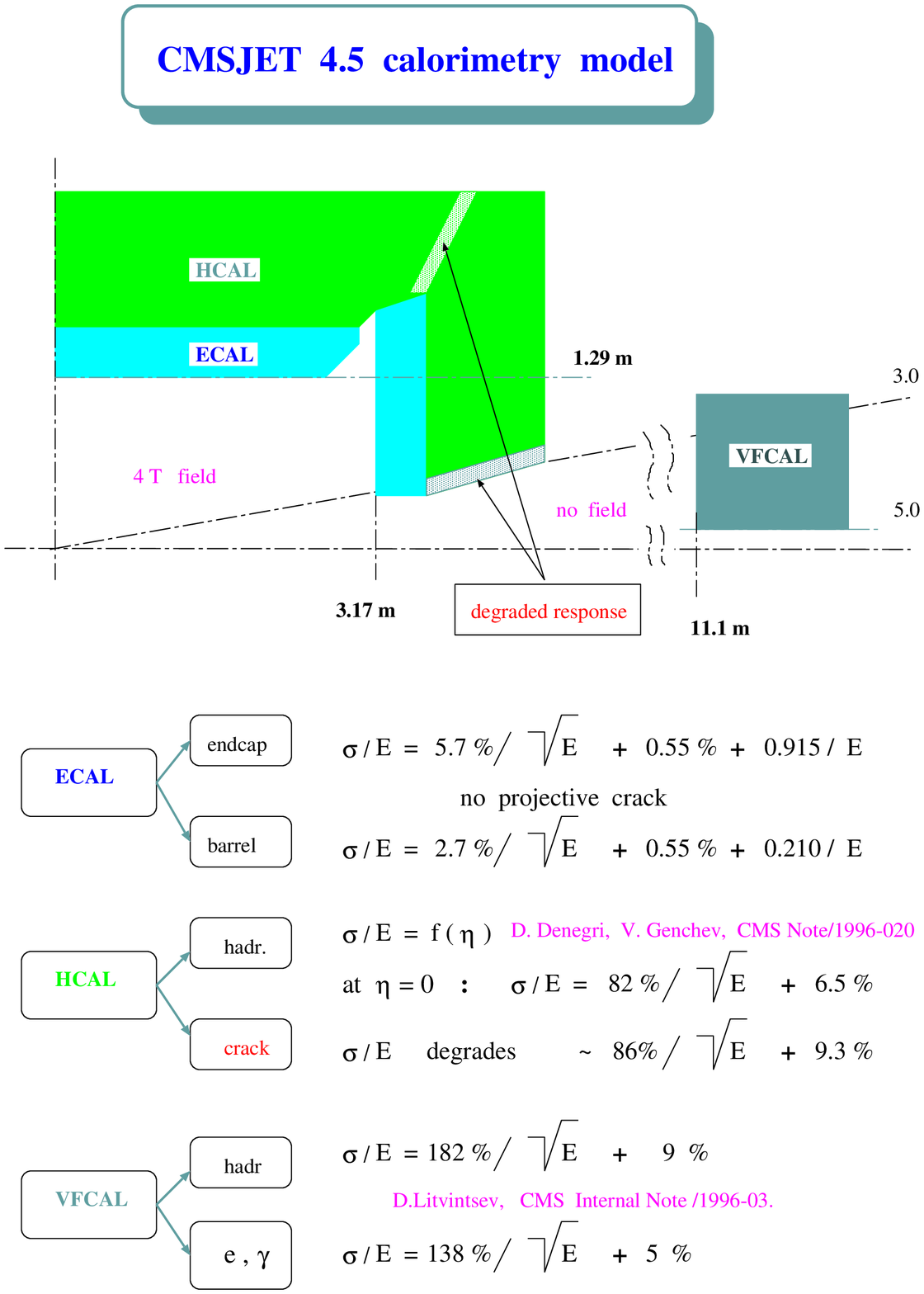}}
\vspace*{5mm} \caption{ Sketch of calorimetry description in CMSJET.}
\end{center}
\end{figure}

It is very difficult to produce a representative sample of QCD jet
background in the low-$\hat{p}_{T}$ range since the cross section is huge and
 we need extreme kinematical fluctuations of this type of background to be
 within the signal selection cuts. Even having spent a couple of CPU 
years and using fast MC we are able to exploit only a tiny fraction
of QCD background of low-$\hat{p}_{T}$ values.
Fortunately, there is a  correlation between $\hat{p}_{T}$ and maximal
produced E$_{T}^{miss}$ value, since the main sources of E$_{T}^{miss}$ in 
QCD events, such as neutrinos form b,c-jets and  E$_{T}^{jet}$ mis-measurement,
strongly correlate with the $\hat{p}_{T}$.
This allows one not to expect
high values of E$_{T}^{miss}$ from low-$\hat{p}_{T}$ QCD events.   

Nevertheless, to be on the safe side, 
we  cannot go confidently  below $\simeq$ 200 GeV
with the cut on  E$_{T}^{miss}$,
where the QCD jet background becomes the dominant contribution and where
our simulations are not yet fully reliable for this type of background,
as it depends on the still evolving estimates of dead areas/volumes due to
services etc.

Initial requirements for all the samples are the following $:$
\begin{itemize}
 \item at least 2 jets with  E$_{T}^{jet}$ $>$ 40 GeV in
 $\mid\eta^{jet}\mid$ $<$ 3
 \item E$_{T}^{miss}$ $>$ 200 GeV
\end{itemize}

In this analysis in general no specific requirements are put on leptons.
If there are isolated muons with p$_{T}^{\mu}$ $>$ 10 GeV within the muon
acceptance,  or isolated electron with
 p$_{T}^{e}$ $>$ 20 GeV within $\mid\eta^{e}\mid$ $<$ 2.4
 in the event, they are also recorded to use them in the subsequent analysis.
The term ``isolated lepton''  here means satisfying simultaneously the
following two  requirements  $:$
\begin{itemize}
 \item no charged particle with p$_{T}$ $>$  2 GeV in a cone R = 0.3 around the
 direction of the lepton,
 \item  $\Sigma$ E$_{T}^{cell}$ in a ``cone ring'' 0.05 $<$ R $<$ 0.3
 around the
lepton impact point has to be less than 10 \% of the lepton transverse energy
\end{itemize}
The electrons are always required to satisfy these isolation criteria due to
identification requirements, whilst muons can be identified even in jets,
so isolation is not mandatory to identify them. Hence muon isolation
requirement can be used to optimize the results.
A factor of ``detection efficiency'' of $\epsilon$=0.9
 is applied for each lepton 
to take into account various inefficiencies of the full pattern recognition.

\section{mSUGRA signal and ``SM background''}

In the following 
the term mSUGRA signal means the sum of all sparticle production processes $:$
$\tilde{g}$, $\tilde{q}$ pair and associated with other sparticles (e.g.
$\tilde{\chi}_{2}^{0}$ and $\tilde{\chi}_{1}^{\pm}$), 
chargino-neutralino pair production etc.  The SM background includes 
processes listed in Tab.2.
All the specific signal final states investigated here mean samples of
events from the total mSUGRA signal passing initial requirements listed in 
section 3 and classified according to presence (or total absence) of a definite
number of isolated electrons and (isolated or not) muons in the final state.
The E$_{T}^{miss}$ signature means that the whole signal sample satisfies
requirements concerning jets and E$_{T}^{miss}$ beyond initial ones, and
 is treated without taking into account possibly identified leptons.
 The 0l signature implies a lepton veto with
the leptonic requirements listed in section 3.
\begin{figure}[hbtp]
\begin{center}
\vspace*{0mm}
\hspace*{0mm}\resizebox{0.85\textwidth}{18cm}
                        {\includegraphics{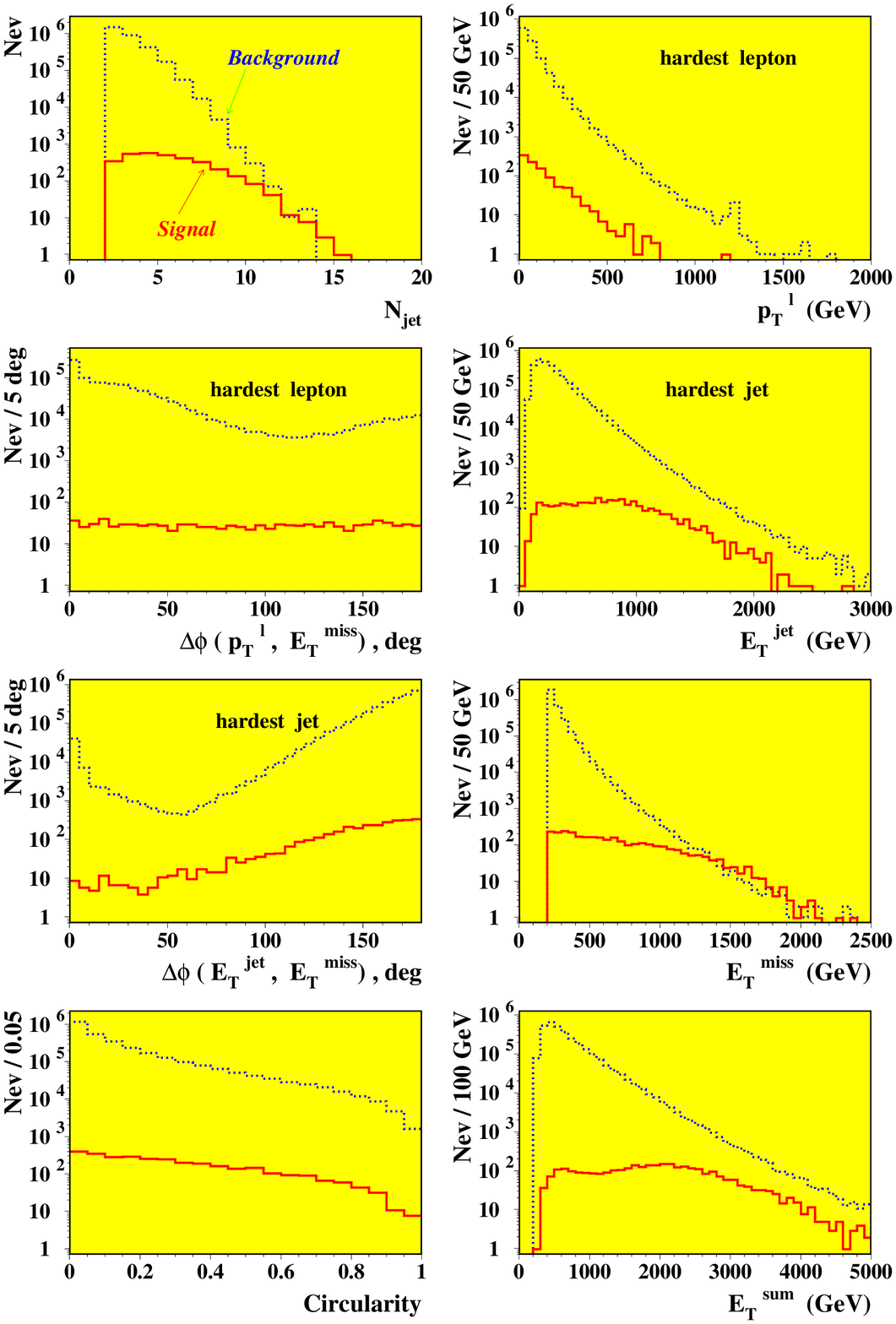}}
 \caption{ Comparison of the mSUGRA signal and SM background,
 with 100 fb$^{-1}$,
in one point of mSUGRA parameter space $:$ m$_0$=1000 GeV, m$_{1/2}$=800 GeV,
  Set 4.
(m$_{\tilde{g}}$$\approx$m$_{\tilde{q}_{L}}$$\approx$ 1900 GeV)
for the E$_T^{miss}$ signature. Initial cuts 
are applied.}
\end{center}
\end{figure}
\begin{figure}[hbtp]
\begin{center}
\vspace*{0mm}
\hspace*{0mm}\resizebox{0.85\textwidth}{18cm}
                        {\includegraphics{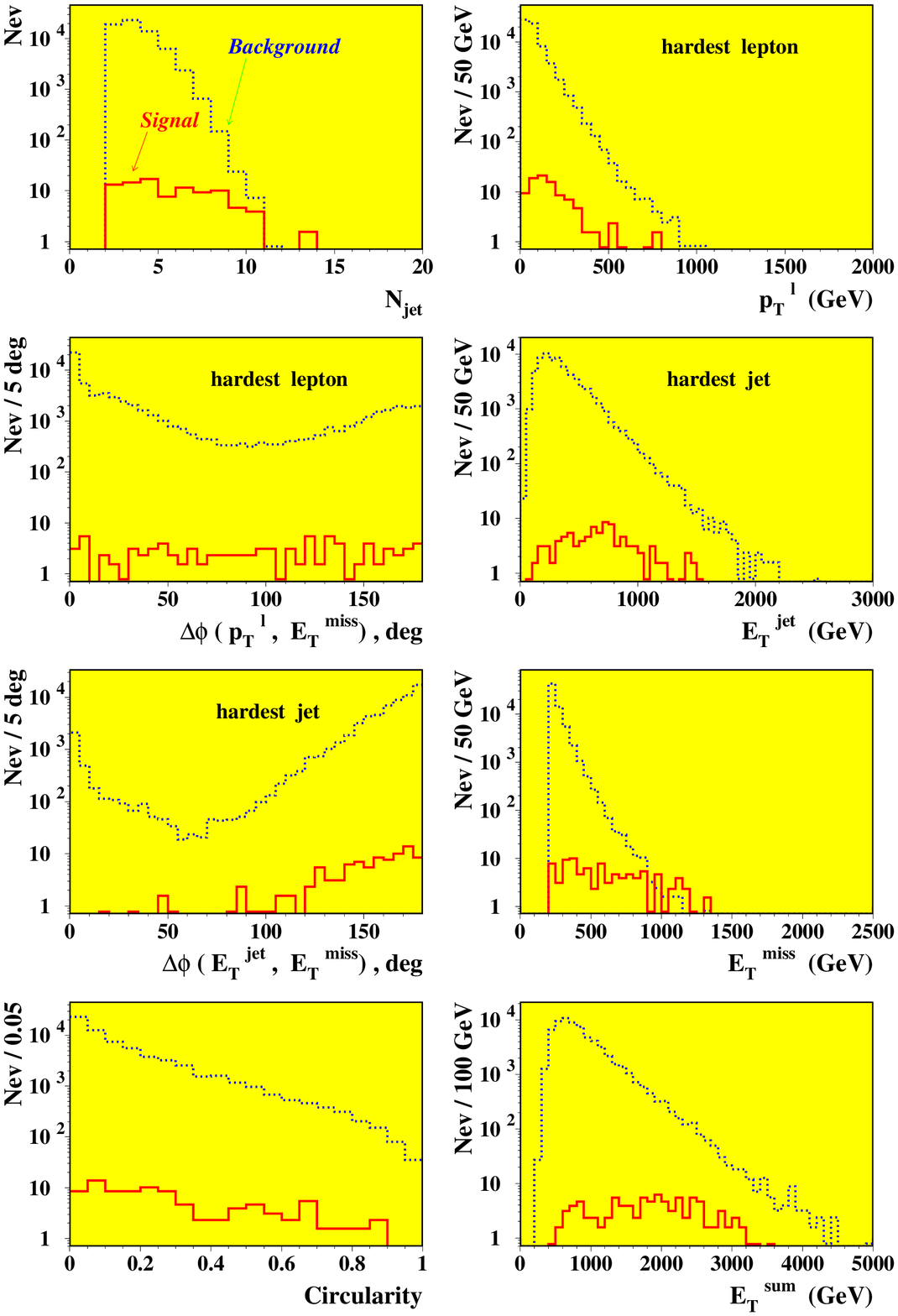}}
 \caption{ Same as Fig.16, except for 2l OS final states.}
\end{center}
\end{figure}
 The 1l signature means presence of a single lepton found in the event,
2l SS - two leptons of same sign, 2l OS - two leptons of opposite sign and
3l - three leptons in the event
with leptons satisfying the basic criteria specified in section 3, and
there are the basic  E$_{T}^{miss}$ and jets requirements.

The kinematics of signal events
is usually  harder than that of SM background
for the interesting regions of maximal reach of squark-gluino masses .    
The cross-section of the background is however higher by orders of magnitude
and high-p$_{T}$ tails of different backgrounds 
can have a kinematics similar to that of the signal.

In Fig.16
 we compare some signal distributions for the  point 
 (m$_{0}$, m$_{1/2}$) = (1000,800) of Set 4, 
 corresponding to  m$_{\tilde{g}} \approx$ m$_{\tilde{q}_{L}} \approx$
 1900 GeV,
m$_{\tilde{\chi}_{1}^{0}}$ = 351 GeV, m$_{\tilde{\chi}_{2}^{0}}$ =
 m$_{\tilde{\chi}_{1}^{\pm}}$ = 668 GeV,
and distributions of the sum of all SM background processes listed in Tab. 2
for the E$_{T}^{miss}$  signature.
Both signal and
background histograms contain only events satisfying first level selection 
criteria.  Only the hardest jet and lepton in the event are shown
 in distributions in
Fig.16.

Fig.17
shows the same comparison at the same mSUGRA point as in
Fig.16,
except for the 2l OS final state signature,
 with non-isolated muons. Both the signal and
background samples are significantly smaller than those in 
Fig.16
(with the same initial cuts).

Since the topology of signal and background events is rather  
similar already  after  first level selection cuts,
the difference in the
angular distributions and circularity is not significant either,  
it is thus not very useful to apply  
cuts on these variables too. The difference in the lepton p$_{T}$ distributions
 is also not very pronounced as  signal leptons are produced in
cascade decays, thus loosing  ``memory'' about the hardness of the original
process. But for extremely high masses of squarks or gluinos ($\sim$ 2 TeV),
there is some difference in the angular and p$_T^{l}$ distributions 
between signal and the total SM background.
For example, the W+jets background contributes significantly in the leftmost
part of the $\Delta\phi(p_{T}^{l},E_{T}^{miss})$ distribution, especially 
in case of the 1l signature with high cuts on E$_{T}^{miss}$ and 
jets E$_{T}^{j}$.
 So cuts on these variables can be useful in these conditions.

One can deduce from Figs.16 and 17 that cuts on the jet multiplicity N$_{j}$
 and E$_{T}^{miss}$
are the most profitable ones for background suppression .
Of course, there is inevitable correlation between variables both in signal 
and background, e.g. an obvious correlation  between
 E$_{T}^{miss}$ and the hardest jet  E$_{T}$ in QCD events, since there
 E$_{T}^{miss}$ is mainly produced by neutrinos from b-jets and/or
 high-E$_{T}^{jet}$ mis-measurement. This can lead to a degradation of
the efficiency of some cuts, if fixed cuts are used. 
It is thus more profitable to have
 adjustable cuts to meet various kinematical conditions
in various domains of mSUGRA parameter space and take into account
difference in topology between various signatures.

Anyway, the cuts have to be justified from the point of view of
the best observability of the signal over expected background, and     
in  all cases, signal observability is based on an excess of
events of a given topology over  known (expected) backgrounds.

\section{Cuts optimization procedure}
\subsection{General considerations}

The chosen criterion of the mSUGRA signal observability is  S $\geq$ 
5 $\cdot$ $\sqrt{S + B}$, where S means number of mSUGRA signal events,
B - number of SM background.
In other words, it can be expressed as signal (= all recorded -
 SM expectations) 
has to be five times larger than the square root of all events kept. 
So cuts have to be adjusted in each probed mSUGRA point in such a way
 that the  observability function
 S / $\sqrt{S + B}$ be maximal.

The set of cuts on selected variables applied in each probed point to both
signal and background to find the best observability is shown in Table 4.
The total number of combinations exceeds 10$^4$, but in practice, only part
of the entire ``cut space'' (1000 - 3000 combinations) is really used
 to optimize the reach for each topology and for each mSUGRA parameter Set.  
The procedure works as follows. All the cut combinations are applied
 independently at each 
probed point of parameter space as well as to the background samples.
The best value of the  observability function is then evaluated in each point,
having summed data over all background channels for each particular cut
combination.  The smooth boundary curve is then found interpolating
between  m$_0$ and m$_{1/2}$ points.
\begin{table}[htb]
  \caption{Sets of cuts.}
  \label{tab:4}
  \begin{center}
    \renewcommand{\arraystretch}{1.2}
    \setlength{\tabcolsep}{2mm}
\begin{tabular}{|c|c|c|} \hline
  Variable(s) &  Values  & Total number \\ 
 \hline \hline
 N$_{j}$        &  2, 3, 4, ..., 10                                    & 9  \\
\hline
 E$_{T}^{miss}$ &  200, 300, 400, ..., 1400 GeV                        & 14 \\
\hline
 E$_{T}^{j1}$ & 40, 150, 300, 400, 500, 600, 700, 800, 900, 1000 GeV& 10 \\
 E$_{T}^{j2}$ & 40, ~80, 200, 200, 300, 300, 400, 400, 500, ~500 GeV&    \\
\hline
 $\Delta\phi$ $(p_{T}^{l},E_{T}^{miss})$ & 0, 20 deg.                  & 2 \\
\hline
$ Circularity $    &  0, ~0.2                                         & 2  \\
\hline
 $\mu$ isolation  &  on, off                                           & 2  \\
\hline
\hline
 total      &     ~                 &  $\approx$  10$^{4}$               \\
\hline
\end{tabular}
\end{center}
\end{table}
\subsection{Numerical examples}

Just to give an idea about orders of magnitude, we show in Table.5 
some numerical examples of best cuts found in a few representative points 
of Set 4.
The points are chosen near the 5 $\sigma$ reach boundary for the  
corresponding experimental final state signature.
\begin{table}[htb]
  \caption{Numerical examples of cuts optimization with integrated luminosity
 of 100 fb$^{-1}$.}
  \label{tab:5}
  \begin{center}
    \setlength{\tabcolsep}{1mm}
\begin{tabular}{|c|c|c|c|c|c|c|c|c|c|c|c|c|c|c|c|} \hline
  & \multicolumn{2}{c}{Point of Set 4}\vline &
 \multicolumn{6}{c}{Cuts values} \vline &
 S  & \multicolumn{5}{c}{B (ev)}\vline & \\
 \cline{2-9} \cline{11-15} 
 Signature & m$_{0}$ & m$_{1/2}$ & N$_{j}$ & E$_{T}^{miss}$ &  E$_{T}^{j1}$,
 E$_{T}^{j2}$  &  $\Delta\phi$ & $Circ.$ & $\mu$-isol. & (ev)  & 
 $t\bar{t}$ & $Wj$ & $Zj$ & $QCD$ & All &  $\frac{S}{\sqrt{S+B}}$  \\
 & (GeV) & (GeV) & & (GeV) & (GeV) & (deg) &  & (on/off) &  & & & & & & \\
\hline
\hline
 E$_{T}^{miss}$ & 500 & 1200 & 2 & 1200 & 900, 600 & 0 & 0 & off & 57.0 &
 4 & 18 & 17.6 & 1 & 40.6 & 5.77 \\  
  \cline{2-16}
          &      1600 & 1000 & 7 & 600  & 600, 300 & 0 & 0 & off & 27.6 &
 1 & 2 & 3.8 & 8.8 & 15.6 & 4.2 \\
\hline
\hline
 1l       &       400 & 1100 & 2 & 900 & 600, 300 & 20 & 0 & on & 31.9 &
 1.8 & 13.2 & 0 & 0 & 15.0 & 4.66 \\
  \cline{2-16}
          &      1000 & 1000 & 4 & 800 & 500, 300 & 20 & 0 & off & 36.0 &
 4.5 & 11.5 & 7.1 & 0 & 23.1 & 4.68 \\
\hline
\hline
 3l       &       400 &  700 & 2 & 300 & 150,  80 &  0 & 0 & on & 41.3 &
   0  & 0.7  & 0 & 0 & 0.7 & 6.37 \\
  \cline{2-16}
          &      1400 &  700 & 2 & 300 & 300, 200 &  0 & 0 & off & 37.7 &
 8 & 2.5 & 15.8 & 0 & 26.3 & 4.72 \\
\hline
\end{tabular}
\end{center}
\end{table}

\section{Results}
\subsection{5 $\sigma$ reach}

Figs.18-21
 show the main results of our study for mSUGRA parameter sets 
given in Tab.1 assuming an integrated luminosity of 100 fb$^{-1}$. 
Fig.22
regroups   
Figs.18-21
 together, without details, just for visual comparison of respective searches.
 The dashed-dotted lines in 
Figs.18-21
 are isomass contours for
squarks ($\tilde{q}$), gluino ($\tilde{g}$)  and lightest scalar Higgs
 ($h$). Numbers in parenthesis denote mass values of corresponding isomass
contour. The neutralino relic density contours 
from ref. \cite{ex_regions}, for mSUGRA domain  m$_{0}$$<$1000 GeV,
 m$_{1/2}$$<$1000 GeV, are also shown in 
Figs.18-21
for $\Omega h^{2}$ = 0.15, 0.4 and 1.0.
Value  $\Omega h^{2}$$>$1 would lead to a Universe age less than 10
billion years old, in contradiction with estimated age of the oldest stars.
The region in between 0.15 and 0.4 is favoured by the Mixed Dark Matter (MDM)
cosmological models.  
\begin{figure}[hbtp]
\begin{center}
\vspace*{0mm}
\hspace*{0mm}\resizebox{0.99\textwidth}{20cm}
                        {\includegraphics{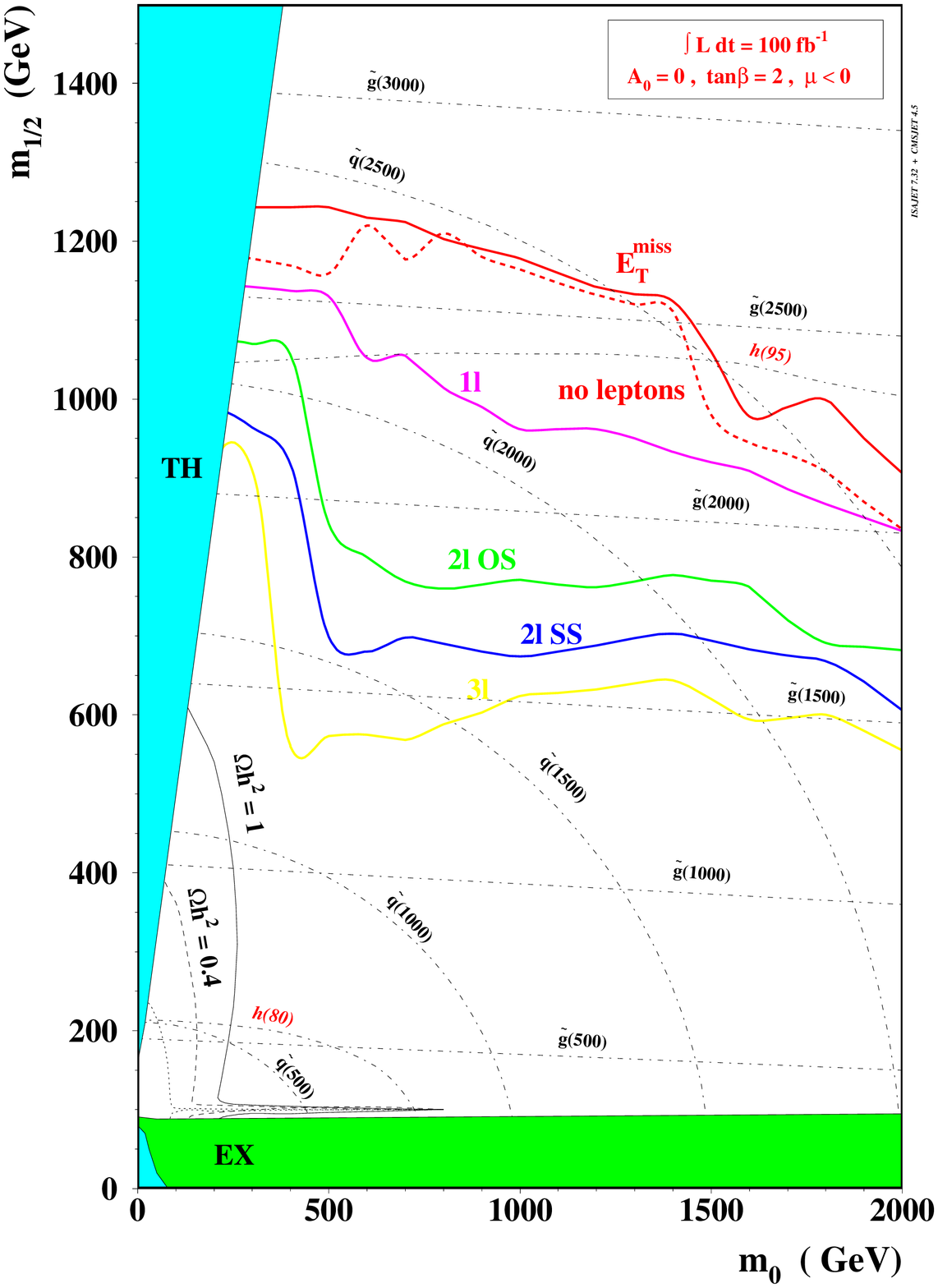}}
 \caption{ 5 sigma reach contours for various final states
 with 100 fb$^{-1}$ for Set 1 (see also comments in text).}
\end{center}
\end{figure}
\begin{figure}[hbtp]
\begin{center}
\vspace*{0mm}
\hspace*{0mm}\resizebox{0.99\textwidth}{20cm}
                        {\includegraphics{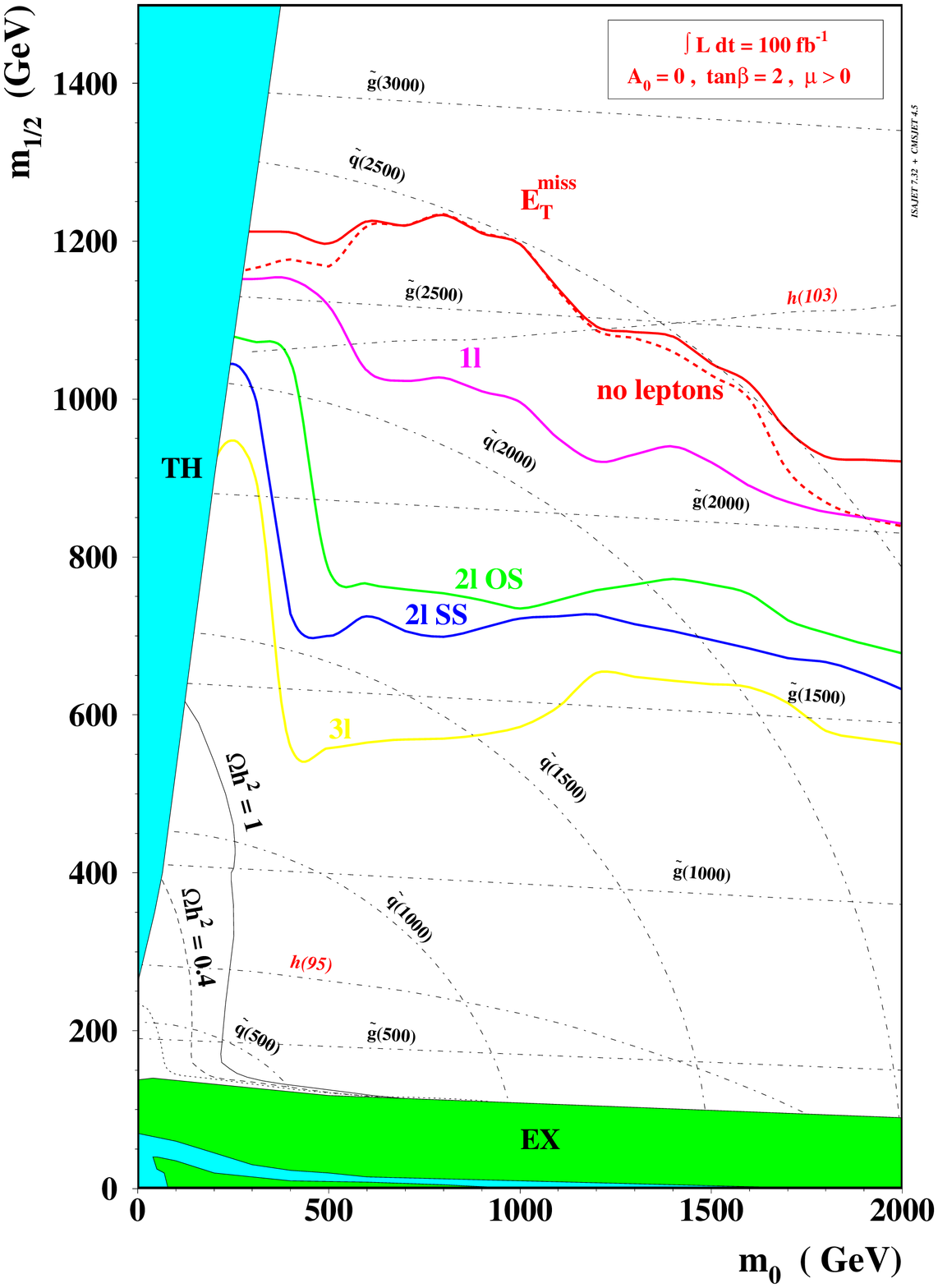}}
 \caption{ Same as Fig.18, but for Set 2.}
\end{center}
\end{figure}
\begin{figure}[hbtp]
\begin{center}
\vspace*{0mm}
\hspace*{0mm}\resizebox{0.99\textwidth}{20cm}
                        {\includegraphics{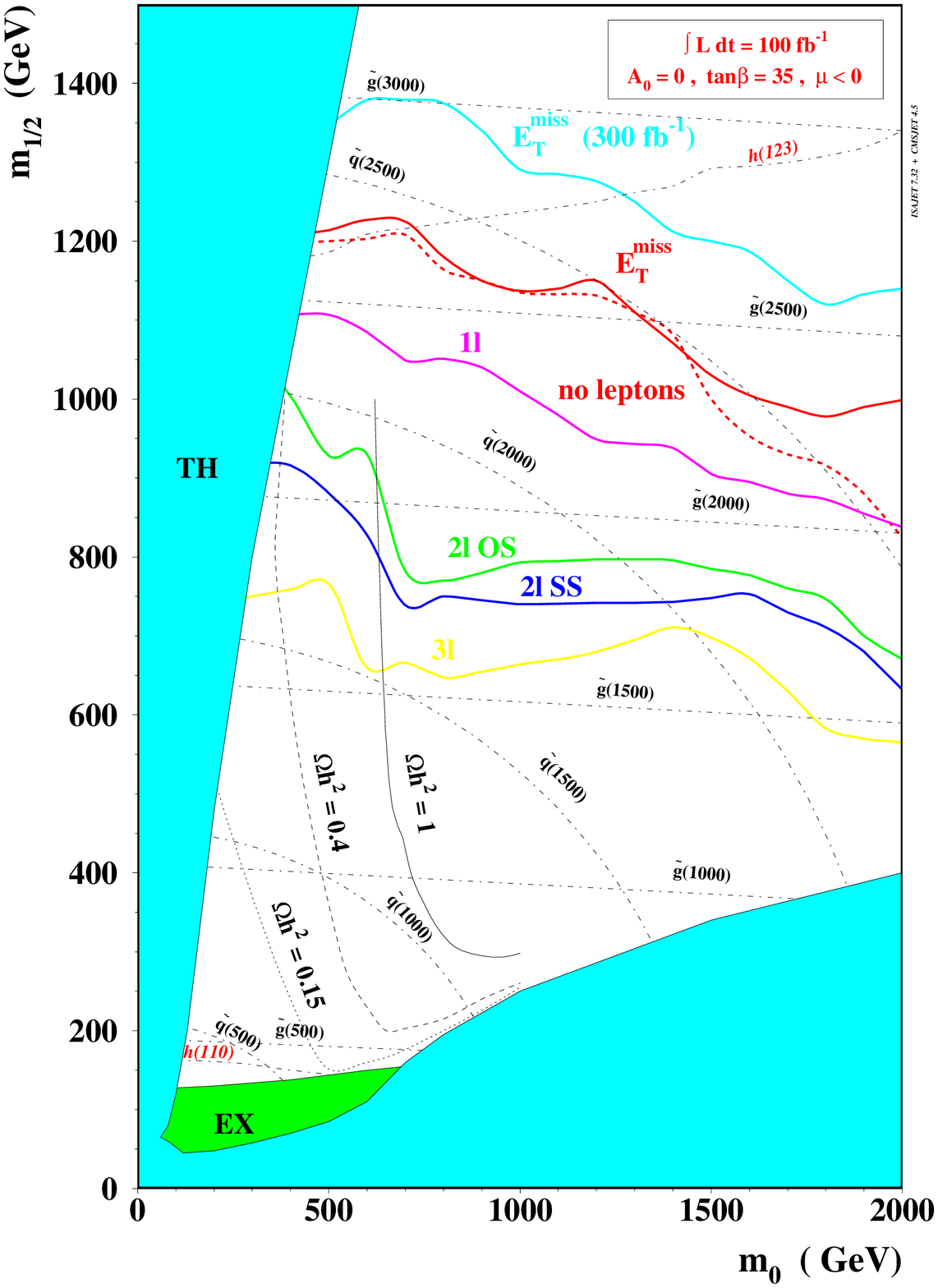}}
 \caption{ Same as Fig.18, but for Set 3.}
\end{center}
\end{figure}
\begin{figure}[hbtp]
\begin{center}
\vspace*{0mm}
\hspace*{0mm}\resizebox{0.99\textwidth}{20cm}
                        {\includegraphics{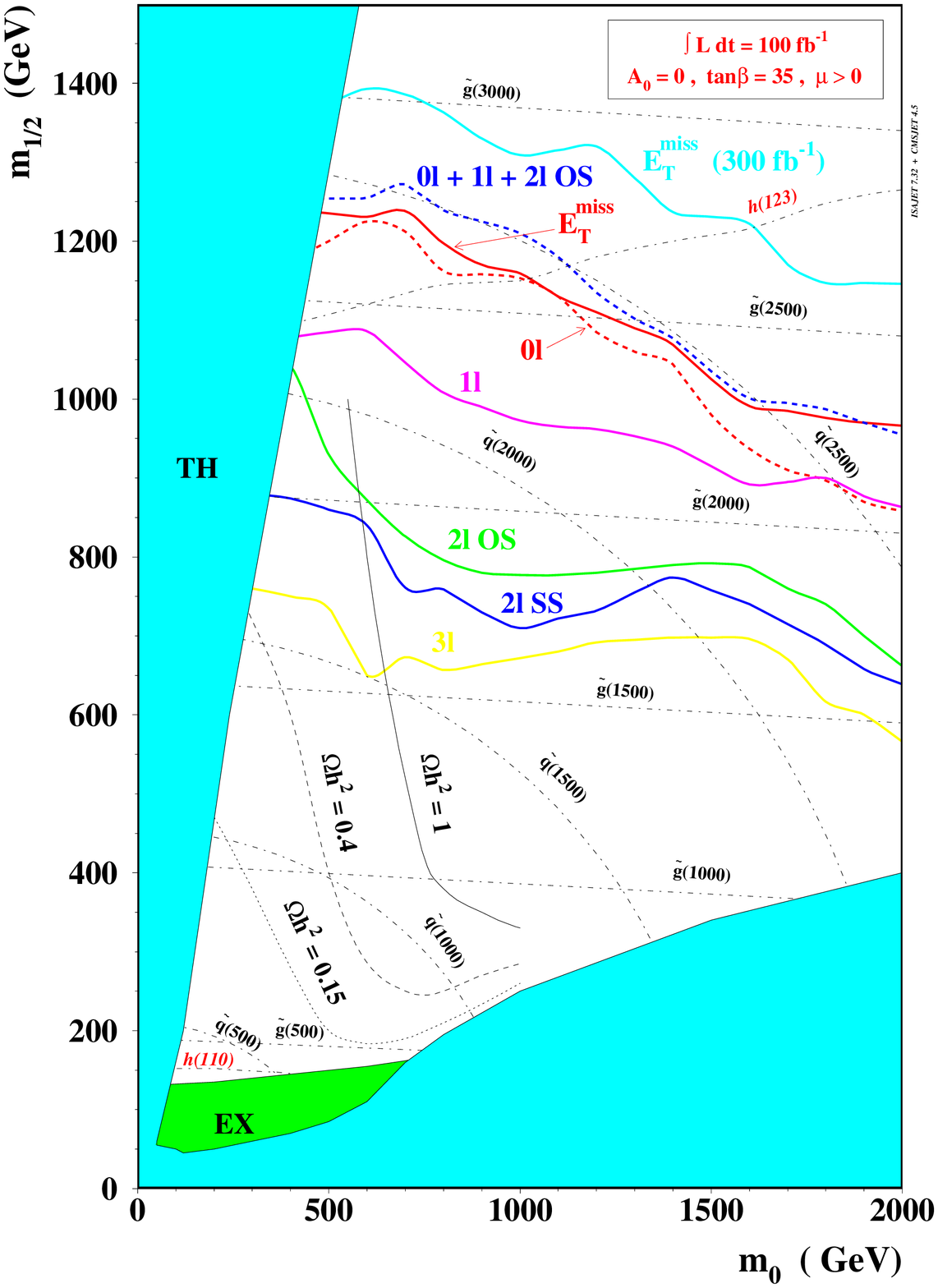}}
 \caption{ Same as Fig.18, but for Set 4.}
\end{center}
\end{figure}
\begin{figure}[hbtp]
\begin{center}
\vspace*{-10mm}
\hspace*{0mm}\resizebox{0.99\textwidth}{20cm}
                        {\includegraphics{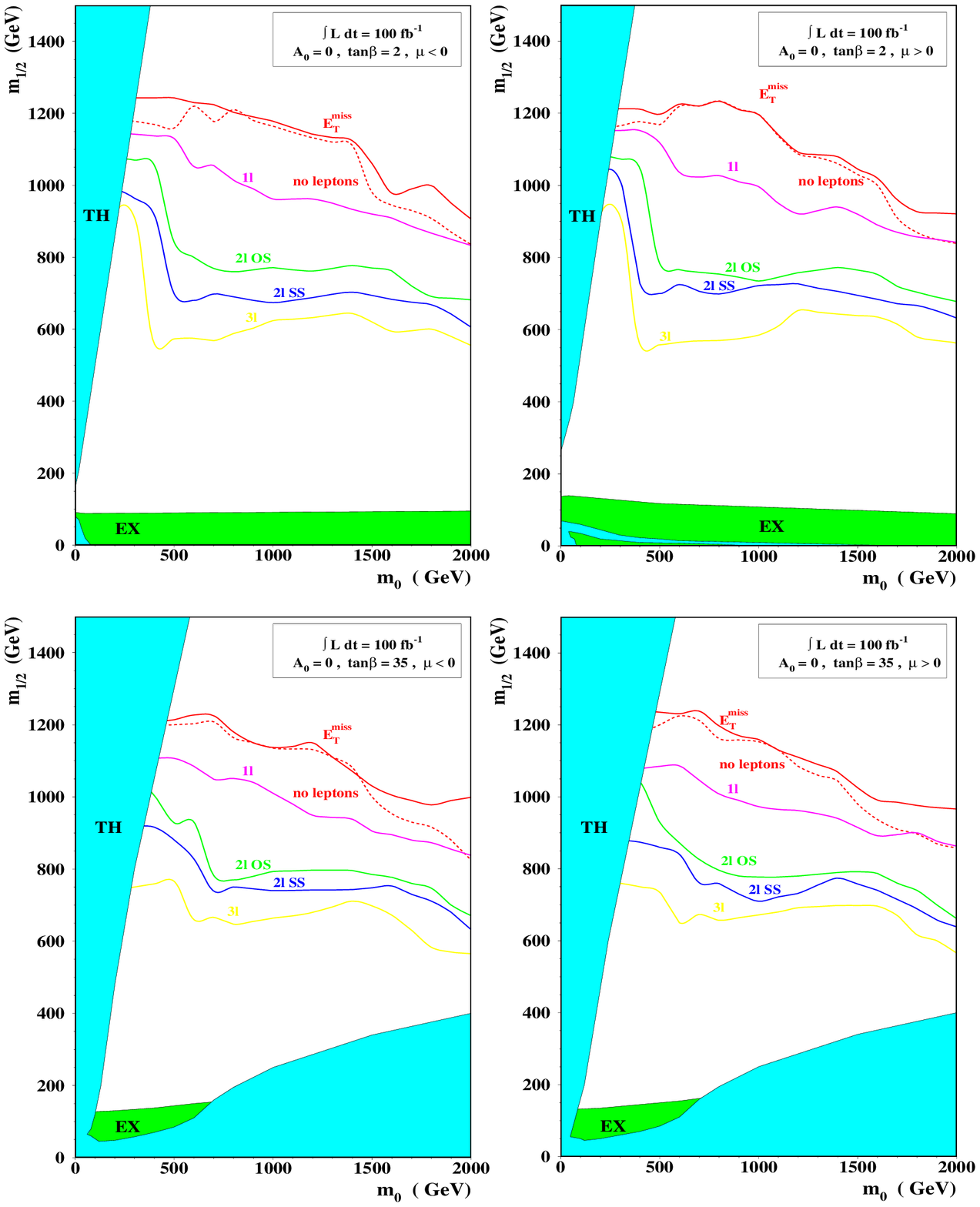}}
\vspace{10mm}
 \caption{ Simplified Figs.18-21 put together.}
\end{center}
\end{figure}

It is  a rather general situation that for all investigated sets of 
mSUGRA parameters the best reach can be obtained with the E$_{T}^{miss}$
 signature. The more leptons required - the smaller reach, as can be seen from 
Figs.18-22. 
For Set 1 one can see that the entire m$_{0}$, m$_{1/2}$ plane  
(Fig.18)
is covered, in principle by LEP II through Higgs searches.
 This is not  the case for Set 2 
(Fig.19).
 Anyway, for low tan$\beta$ cosmologically 
interesting regions will be definitely covered by LEP II or LHC reach.  
I does not seem so evident for high values of tan$\beta$ 
(Figs.20 and 21),
where  the calculations of  $\Omega h^{2}$=1 contours  
available are limited to  m$_{1/2}$ $\leq$ 1000 GeV. But again,
the cosmologically
preferred region  $\Omega h^{2}$$<$0.4 seems to be entirely within the reach
 of CMS. In both
Figs.20 and 21
 we also show our calculations for  the E$_{T}^{miss}$
signature reach for an integrated luminosity of 300 fb$^{-1}$,
 trying to estimate
the ultimate CMS reach.
The Higgs contours in
 Figs.20 and 21 
show that most of the
 m$_{0}$, m$_{1/2}$ planes is out of reach for LEP II.

It is worth noticing that the cumulative reach of several signatures, like
0l + 1l + 2l + 3l + ... (in descending order of contribution) 
can be even better than the most promising single  
 E$_{T}^{miss}$ signature. We show this with the cumulative
0l + 1l + 2l OS signatures curve (main leptonic signatures) in 
Fig.21.
Despite the fact that this curve does is not obtained with
 the optimally adjusted cuts 
(each signature was optimized separately to have the best significance
 and then signal and
background values were summed up for all three signatures), one can see that  
the reach obtained is better than that of the E$_{T}^{miss}$ signature. 

Here we do not consider the limitations on the mSUGRA parameter
 space imposed by the $b$ $\rightarrow$ $s\gamma$ calculations 
\cite{bsgam} based on the data from CLEO \cite{CLEO}. These calculations
exclude at 95 \% CL the part of  m$_{0}$, m$_{1/2}$  plane
 approximately below squark isomass contour of 1600 GeV for Set 3 and
 below a similar contour of $\approx$ 600 GeV for Set 4 respectively.
For low tan$\beta$ (Sets 1 and 2) the mSUGRA parameter space domain
 excluded this way is rather small \cite{bsgam_lowtan}. 

\subsection{Effect of the muon isolation requirement on the reach}

\begin{figure}[hbtp]
\begin{center}
\vspace*{-20mm}
\hspace*{0mm}\resizebox{0.99\textwidth}{21cm}
                        {\includegraphics{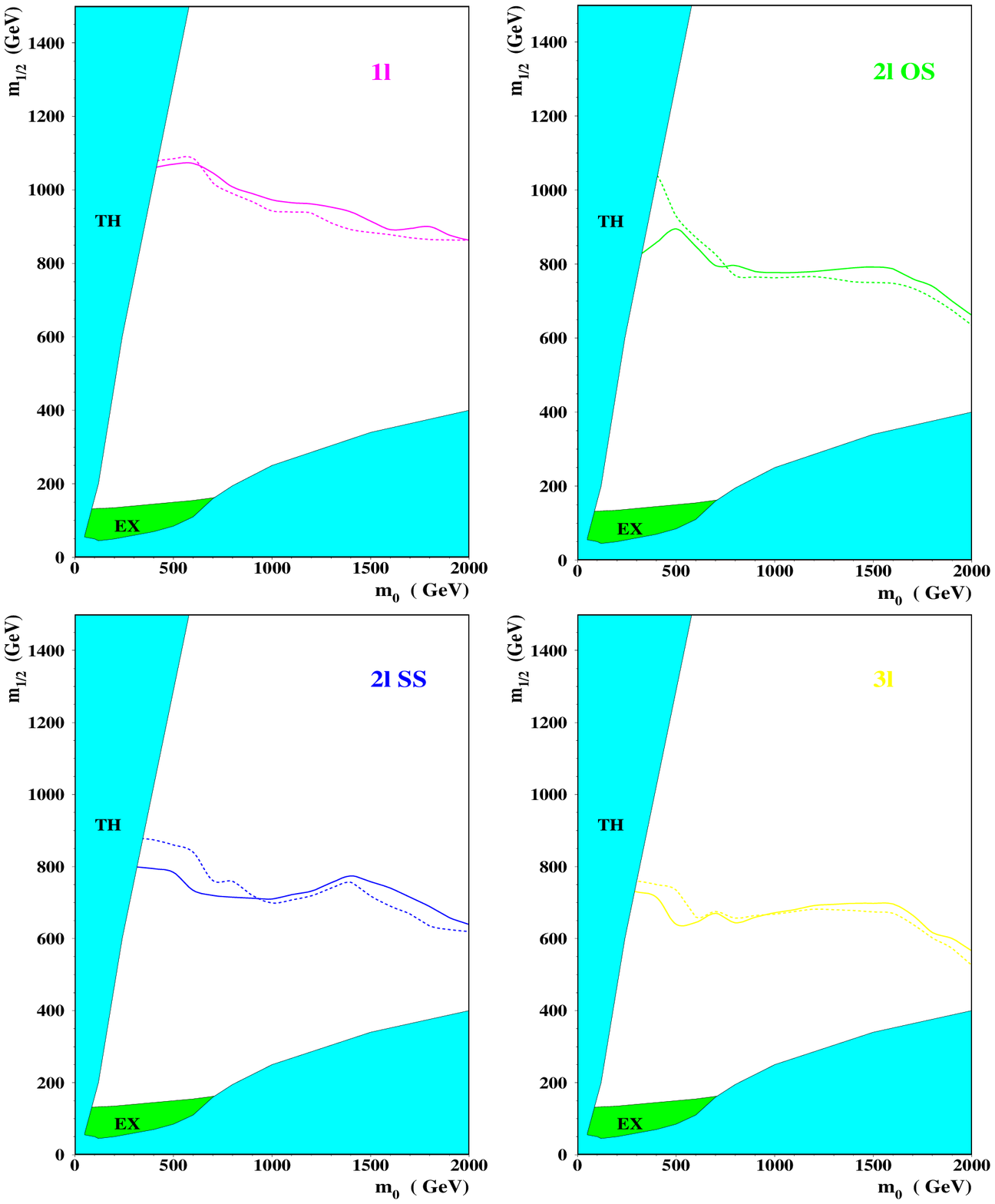}}
\vspace{10mm}
 \caption{ Effects of the muon isolation requirement on the 5$\sigma$ contours
for Set 4.
The reach without a muon isolation requirement is shown by a solid line,
the dashed line denotes the reach with the muon isolation requirement applied.}
\end{center}
\end{figure}
In our study we always require that the isolation criterion mentioned
 in section 3 be fulfilled by electrons, so as to be identified in the CMS
detector. On the contrary the muon isolation requirement
 is a parameter in the
cuts optimization procedure described in section 5. 
This is so as a muon can be identified even when it is embedded in a jet.
Fig.23 
illustrates the effect of the muon isolation requirement on the reach
contours for various multilepton signatures for mSUGRA parameter Set 4
with 100 fb$^{-1}$. 
One can see that for low values of m$_{0}$ it is profitable to apply
muon isolation (dashed line), whilst for high m$_{0}$ values some small
gain can be obtained by not applying the muon isolation criterion (solid line).
This is due to some difference in the origin of leptons in these two domains
of m$_{0}$. As it was discussed in section 2
 ( can also be seen in 
Figs.7 and 8),
the sources of isolated leptons in the low-m$_{0}$ region of Set 4
are mainly $\tau$ and W abundantly produced in the cascade decays
 of squarks and gluinos.
In the high-m$_{0}$ region of Set 4 the situation 
changes slightly $:$ W-bosons are still present in cascades of 
strongly interacting sparticles, but increasing average number of b-jets
(e.g. $\tilde{\chi}_{2}^{0}$ almost completely decay into
$\tilde{\chi}_{1}^{0} h$, and Higgs in turn has branching into $b\bar{b}$
of  $\approx$ 80-90 \%)
contributes significantly to the increase of non-isolated muon production.

\subsection{Stability of results versus variations of 
            signal and background cross sections}

As can be seen from the numerical examples shown in Tab.5, in the vicinity
of 5 $\sigma$ reach boundary the S/B ratio is generally $>$ 1.
 So one can expect that 
the stability of the 5 $\sigma$ reach in terms of 
 S / $\sqrt{S + B}$ depends largely on the variations of the signal rather 
than on the background. We are aware however that PYTHIA can underestimate
the W/Z + jets cross section, especially for multijet events.
Fig.24
 shows the band of the 5  $\sigma$ reach for  E$_T^{miss}$ and 
 2l SS signatures for Set 4 induced by varying the mSUGRA signal cross section
 within $\pm$ 30 \% around the nominal value given by ISAJET. In
Fig.25
one can see the uncertainty band similar to that in
Fig.24,
but now the SM background cross section is varied
by a factor of 2. The width of the
 band for the 2l SS signature in
Fig.25
 is rather small and is
 of the order of pure statistical uncertainties.

\subsection{Effect of event pile-up on the results}

Adding event pile-up to the SM background at the generation stage
causes a few times larger CPU consumption by the CMSJET package
than without pile-up,
since the number of fired cells (crystals, towers) to be processed increases 
significantly. Besides, in the frame of the mentioned package, only a
limited sample of pile-up events (10$^{3}$-10$^{4}$ bunch crossings)
 can be used as an external 
input file due to the large scale (hundreds of Mb) of pile-up hits file
(similar to that of CMSIM one). The 
limited size of pile-up sample means inevitably some bias in 
the data. These are the two main reasons why the SM background
files were produced without pile-up admixture. 
\begin{myfig2}{hbtp}
\vspace*{-5mm}
\hspace*{-5mm}\resizebox{.45\textwidth}{9cm}
                        {\includegraphics{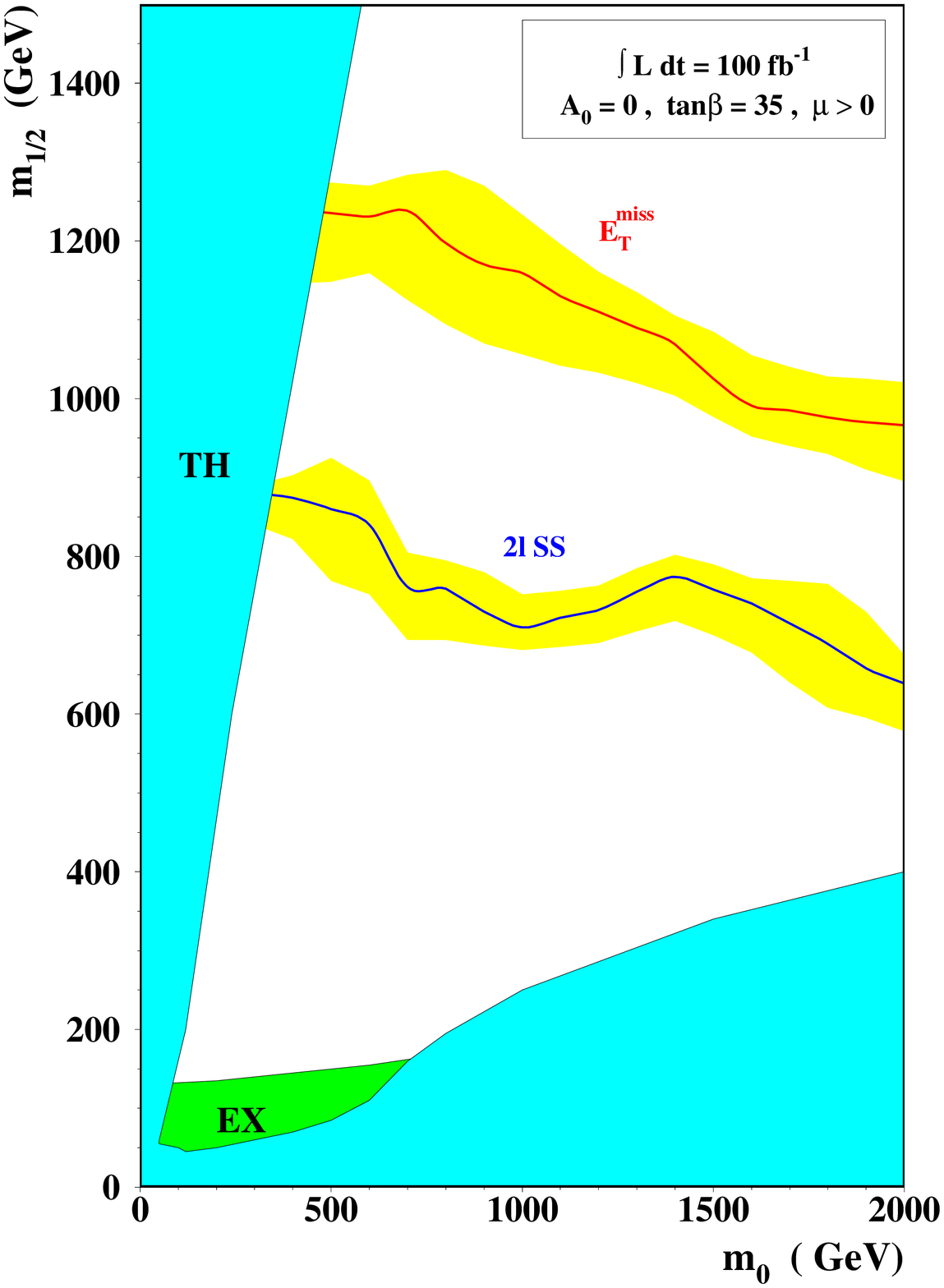}} & ~ &
\vspace*{-5mm}
\hspace*{-5mm}\resizebox{.45\textwidth}{9cm}
                        {\includegraphics{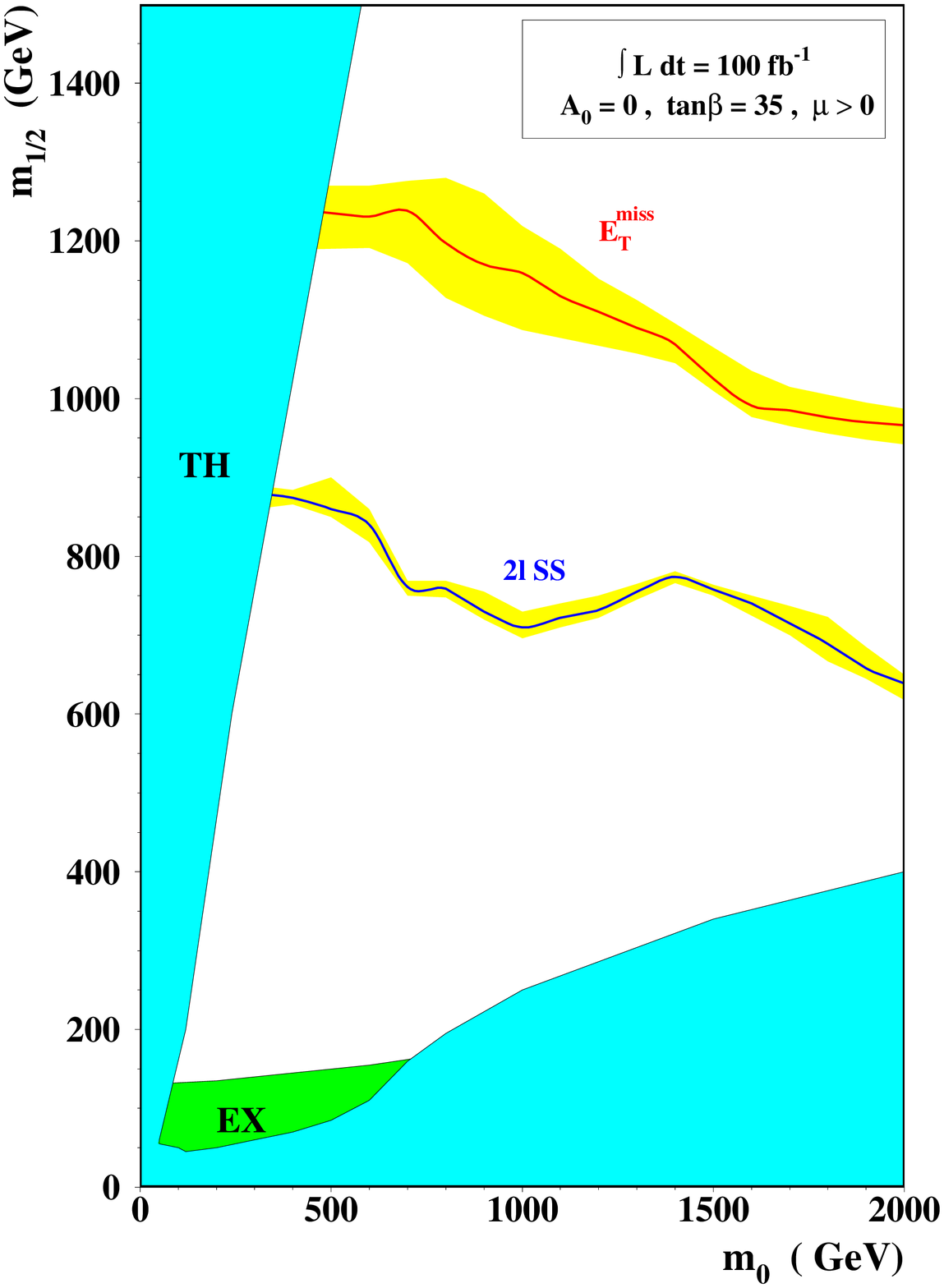}} \\
\vspace*{-5mm}
 \caption{ The band of variation of the reach contour for E$_T^{miss}$ and
  2l SS  signatures with 100 fb$^{-1}$ for a $\pm$ 30 \% variation of 
           mSUGRA cross-sections for Set 4.
 The upper band edge corresponds to an increase
  of the mSUGRA cross section by 30 \%, the lower one - to the same decrease.}
    &~&
\vspace*{-5mm}
\caption{ Same as Fig.24, except for variation of SM cross-sections by a factor
          of 2. The upper band edge corresponds to SM cross-sections decreased
          by a factor of 2, the lower one - to the same increase.}
\end{myfig2}

It is evident that in heavy event pile-up conditions one can
 expect some changes in 
both kinematical distributions (increase of the mean jet number, 
degradation of E$_T^{miss}$ resolution etc.)
 and deterioration of the lepton isolation.
To estimate the effect of pile-up on the signal and background distributions,
we compare some of the main distributions with in the presence of pile-up
and without it. Pile-up is taken from PYTHIA's MSEL=2 with the 
$<$25$>$ interactions per bunch crossing.
\begin{figure}[hbtp]
\begin{center}
\vspace*{-10mm}
\hspace*{0mm}\resizebox{0.7\textwidth}{10cm}
                        {\includegraphics{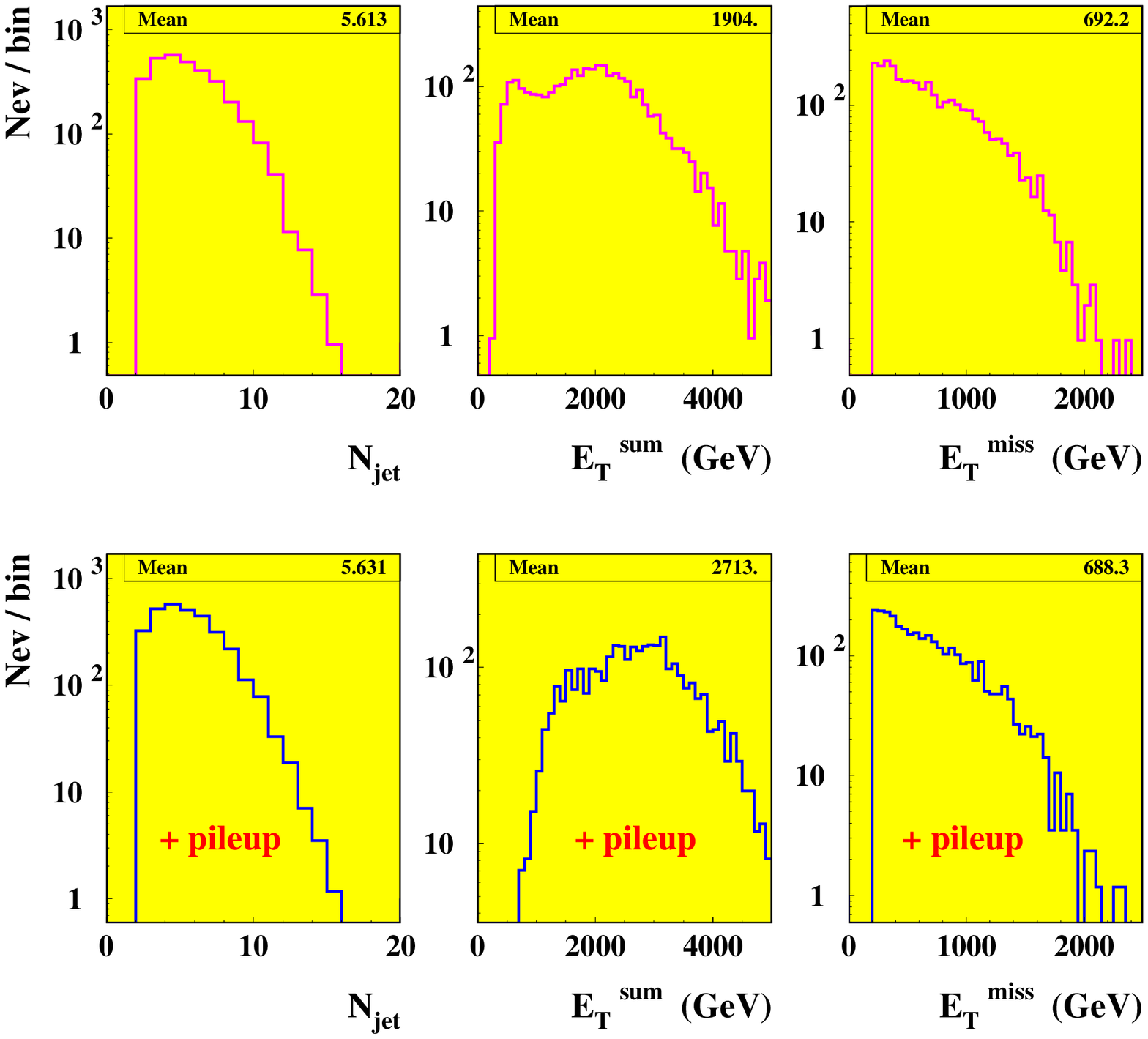}}
 \caption{ Effect of event pile-up on the mSUGRA distributions at one point
of parameter space $:$ m$_0$=1000 GeV, m$_{1/2}$=800 GeV 
 for Set 4 (same point as in Figs.15,16).}
\end{center}
\end{figure}
\begin{figure}[hbtp]
\begin{center}
\vspace*{-5mm}
\hspace*{10mm}\resizebox{0.75\textwidth}{10cm}
                        {\includegraphics{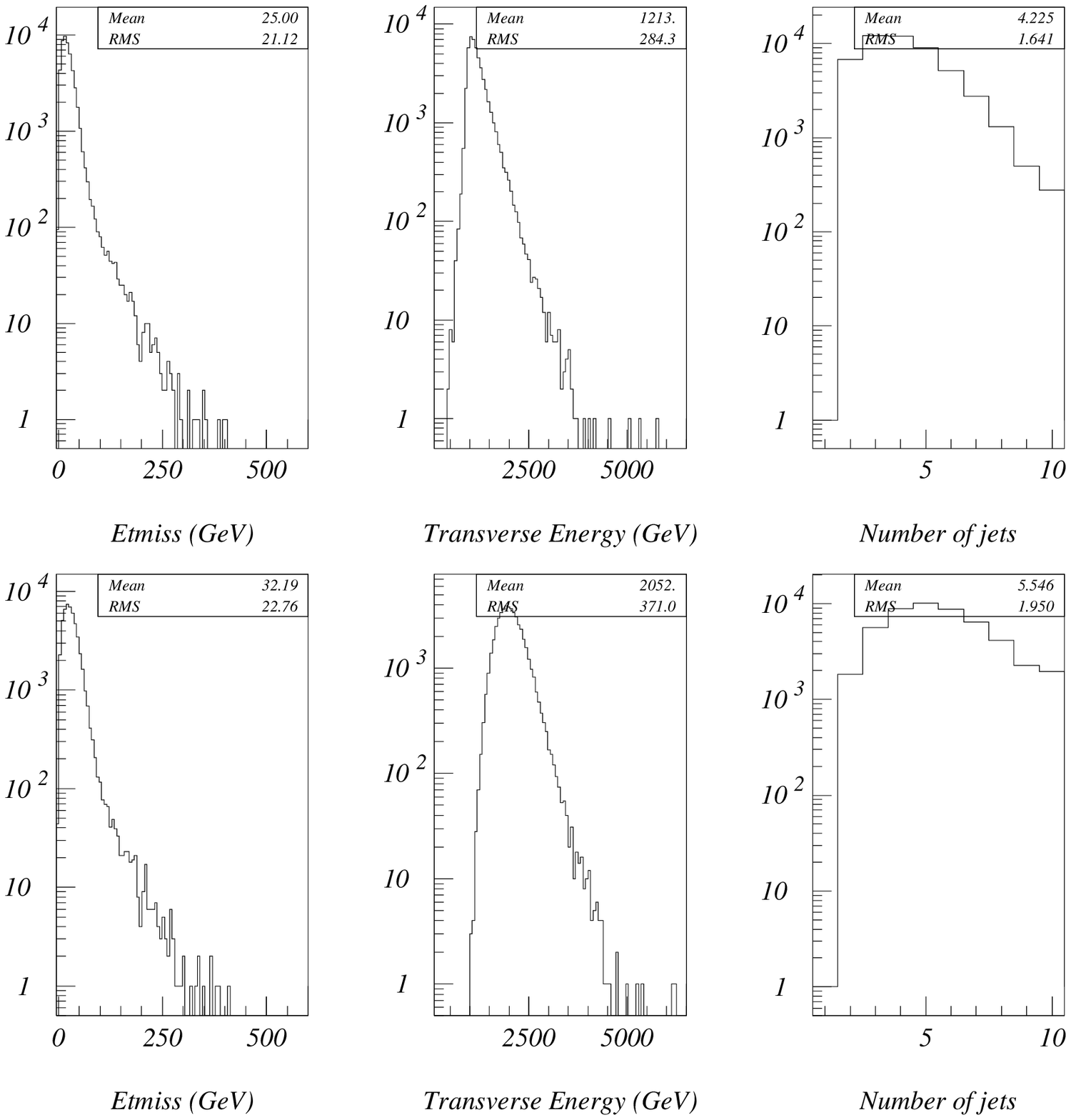}}
 \caption{ Effect of event pile-up on some QCD distributions 
 (400$<\hat{p}_{T}<$800 GeV). The upper row of figures is produced without
pile-up, the lower ones - with pile-up added.} 
\end{center}
\end{figure}

Fig.26
 shows the effect of pile-up on distributions on the number of jets,
the summed  E$_T$ flow through detector and  
the E$_T^{miss}$ at some representative point of parameter space.
 One can see a
rather insignificant effect of pile-up on the distributions presented,
except for the increase of the transverse energy by $\approx$ 800 GeV,
 this is just the quantitative characteristic of the
 pile-up itself, the same increase is expected for any kind of events.
Fig.27
 shows effects of pile-up on the same distributions as in 
Fig.26,
but for the QCD background with 400$<\hat{p}_{T}<$800 GeV.
Here we see again an increase in the transverse energy flow 
by similar value of $\approx$ 800 GeV.
There is also a non-negligible increase in the average number
 of jets (over E$_{T}^{jet}$ = 40 GeV threshold) in QCD events, what is
not observed in  
Fig.26
 for the probed signal point. The reason is rather
simple, the indirect evidence can be derived from
 Fig.16,
 where the hardest
jet distributions for signal and background are compared. One can see that 
total background distribution even for hardest jet falls down very quickly. 
The increase of 
the average number of the jets passing the cut in the QCD case is due to
an exponential fall-off of the ``softest'' jet (originating
mainly from initial and final state radiation) distribution,  
which shifts slightly to the right due to pile-up and giving rise to a
significant number of additional jets in the event. This is not the case for
the chosen signal point with
 m$_{\tilde{g}}$$\approx$m$_{\tilde{q}}$$\approx$1900
 GeV and having all the jets with significant E$_{T}^{jet}$.   

Another possible effect of the pile-up impact on the mass reach could be due to
the deterioration of leptonic isolation. We estimated the average
(over p$_{T}^{l}$,  $|\eta_{l}|$ etc.) loss of
isolated leptons as 10-15 \% per lepton due to pile-up. These losses
are already (at least in part) taken into account by a ``detection
 efficiency'' factor of 0.9 mentioned in section 3.
To estimate the total (cumulative effect) of the pile-up on the
previously calculated 5$\sigma$ contours, we admixed pile-up to all 
 m$_0$, m$_{1/2}$ points of Set 4 and re-evaluated 
the reach contours using the technique
 described in section 5 for the three final states
$:$ E$_T^{miss}$, 1l and 2l OS. The results are shown in 
Fig.28.
 One can see some significant effects of pile-up only in case of 
2l OS final states for low values of m$_0$, where the 
initial isolation of leptons
(originating mainly from $\tau$ and W) can be spoiled by pile-up.
It is worth reminding that both cases of isolated or non-isolated muons
are treated to evaluate the best observability of the signal and 
cuts optimization procedure eventually allows some recovery 
in case of a non-dramatic signal losses.
\begin{figure}[hbtp]
\begin{center}
\vspace*{0mm}
\hspace*{0mm}\resizebox{0.6\textwidth}{12cm}
                        {\includegraphics{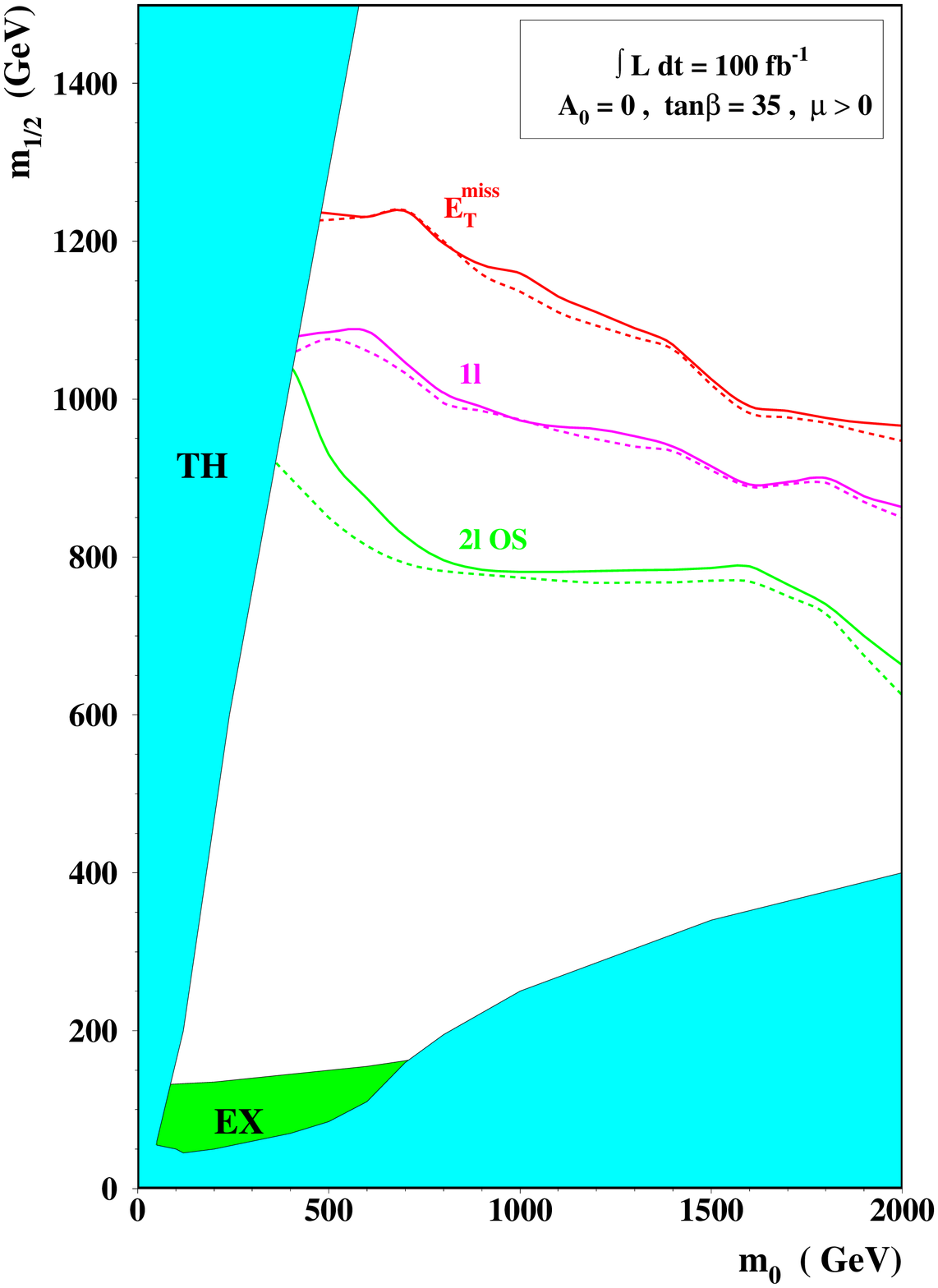}}
 \caption{ Effects of pile-up on the reach contours with
            100 fb$^{-1}$ for Set 4.
The solid line denotes contours obtained without pile-up, dashed line refers to
pile-up added to signal samples.}
\end{center}
\end{figure}

\section{Conclusions}

 The main conclusions of our study are the following  $:$ within the  
SUGRA model investigated SUSY
 would be detectable through
 an excess of events over SM expectations 
up to masses m$_{\tilde{q}}$ $\sim$
m$_{\tilde{g}}$ $\sim$ 2.5 TeV 
with 100 fb$^{-1}$. 
 This means that the entire plausible domain of EW-SUSY 
parameter space for most probable values of tan$\beta$
 can be probed.
Furthermore, the S/B ratios are $>$ 1 everywhere in the reachable domain of 
parameter 
space (with the appropriate cuts) thus allowing a study of kinematics of 
$\tilde{q}$, $\tilde{g}$ production and obtaining information  
on their masses.
 The cosmologically interesting region $\Omega h^{2} \leq$ 1, and even more 
so  the preferred region  0.15 $\leq$ $\Omega h^{2} \leq$ 0.4,
 can be entirely  probed.

\section{Acknowledgements}

We would like to thank prof. Daniel Huss for the support of this work
and prof. Daniel Denegri for his advices and fruitful discussions.
We also thank Alexandre Nikitenko for help in CMSIM drawings of mSUGRA events.


\vspace{15mm}

\begin{thebibliography}{99}
\bibitem{reach_papers}  H.~Baer, C.-H.~Chen, F.~Paige and X.~Tata,
  Phys.Rev. {\bf D52}, 2746 (1995); Phys.Rev. {\bf D53}, 6241 (1996). 
\bibitem{CMS} CMS collaboration, {\em Technical Proposal}, CERN/LHCC 94-38.
\bibitem{mynote} S.~Abdullin, CMS TN/96-095,\\
 S.~Abdullin, \v{Z}.~Antunovi\'{c} and M.~D\v{z}elalija,
  CMS Note 1997/016.
\bibitem{msugra} H.~Baer, C.-H.~Chen, R.~Munroe, F.~Paige and X.~Tata,
  Phys.Rev. {\bf D51}, 1046 (1995).
\bibitem{large_tan}  H.Baer {\em et al.}, FSU-HEP-980204, hep-ph/9802441.
\bibitem{ex_regions} H.~Baer and M.Brhlik, Phys.Rev. {\bf D57,} 567 (1998). 
\bibitem{Boer} W. de Boer, IEKP-KA/97-03, hep-ph/9705309.
\bibitem{ISAJET} F.~Paige and S.~Protopopescu, in {\em Supercollider Physics},
  p. 41, ed. D.~Soper (World Scientific, 1986); H.~Baer, F.~Paige,
  S.~Protopopescu and X.~Tata, in {\em Proceedings of the Workshop on Physics
  at Current Accelerators and Supercolliders}, ed. J.~Hewett, A.~White and
  D.~Zeppenfeld (Argonne National Laboratory, 1993).
\bibitem{CMSJET} S.~Abdullin, A.~Khanov and N.~Stepanov, CMS TN/94-180 \\
(see also /afs/cern.ch/user/a/abdullin/public/cmsjet/4.5/cmsjet4504\_guide.ps).
\bibitem{CMSIM} C.~Charlot {\em et al.}, CMS TN/93-63.
\bibitem{pictures} http://cmsdoc.cern.ch/cmsim/pictures/cmsim\_events.html .
\bibitem{PYTHIA} T.~Sj\"{o}strand,
  {\em Computer Physics Commun.} 39 (1986) 347;
  T.~Sj\"{o}strand and M.~Bengtsson,
  {\em Computer Physics Commun.} 43 (1987) 367;
  H.U.~Bengtsson and T.~Sj\"{o}strand,
  {\em Computer Physics Commun.} 46 (1987) 43;
  T.~Sj\"{o}strand, CERN-TH.7112/93.
\bibitem{bsgam} H.~Baer, M.~Brhlik, D.~Casta$\tilde{n}$o and X.~Tata,
                Phys.Rev. {\bf D58}, 015007 (1998).
\bibitem{CLEO} M.S.~Alam {\em et al.}, (CLEO collaboration),
                  Phys.Rev.Lett {\bf D74}, 2885 (1995). 
\bibitem{bsgam_lowtan} H.~Baer, M.~Brhlik, Phys.Rev. {\bf D55}, 3201 (1997).  

\end{thebibliography}
\end{document}